\documentclass[english,aps,twocolumn,showpacs,amsmath,amssymb,prx,footinbib]{revtex4-2}
\usepackage[T1]{fontenc}
\usepackage[latin9]{inputenc}
\usepackage{amsmath}
\usepackage{amssymb}
\usepackage{graphicx}

\makeatletter
\usepackage{graphicx}
\usepackage{xcolor}
\usepackage{amsmath}
\usepackage{psfrag}
\usepackage[abs]{overpic}
\usepackage{enumitem}
\usepackage{natbib}
\usepackage{float}
\usepackage{verbatim}
\usepackage{soul}

\makeatother

\usepackage{babel}
\begin{document}
\title{Cooper-pair transistor as a minimal topological quantum circuit}
\author{T. Herrig}
\email{t.herrig@fz-juelich.de}
\author{R.-P. Riwar}
\affiliation{Peter Gr\"unberg Institute, Theoretical Nanoelectronics, Forschungszentrum
J\"ulich, D-52425 J\"ulich, Germany}
\begin{abstract}
The outlook of protected quantum computing spurred enormous progress in the search for topological materials, sustaining a continued race to find the most experimentally feasible platform. Here, we show that one of the simplest quantum circuits, the Cooper-pair transistor, exhibits a nontrivial Chern number which has not yet been discussed, in spite of the exhaustive existing literature.
Surprisingly, the resulting quantized current response is robust with respect to a large number of external perturbations, most notably low-frequency charge noise and quasiparticle poisoning. Moreover, the fact that the higher bands experience crossings with higher topological charges leads to all the bands having the same Chern number, such that there is no restriction to stay close to the ground state. Remaining small perturbations are investigated based on a generic master equation approach.
Finally, we discuss a feasible protocol to measure the quantized current.
\end{abstract}
\maketitle

\section{Introduction}

Topological phases are an important research topic in condensed matter
physics~\citep{Bansil:2016aa} most notably with the goal to realize
inherently protected quantum computing~\citep{Sarma:2015aa}. The
most common approach is to search for topological transitions in the
band structure of the materials themselves~\citep{Chiu:2016aa,Hasan:2010aa,Qi:2011aa},
such as in topological insulators~\citep{Kane2005Graphene,Kane2005TopologicalOrder,Bernevig:2006aa,Fu:2007aa,Murakami:2007aa},
Chern insulators~\citep{Hatsugai:1993aa,Thonhauser:2006aa,Prodan:2010aa},
Weyl and Dirac semimetals exhibiting Fermi-arc surface states~\citep{Wan:2011aa,Burkov:2011aa,Hosur:2013aa,Lv:2015aa,Bernevig:2018aa,Armitage:2018aa},
or topological superconductors hosting Majorana fermions~\citep{Sato:2017aa,Kitaev2001,Fu:2008aa,Alicea:2012aa,Meng:2012aa,Badiane:2013aa,Yang:2014aa,Bednik:2015aa,Lutchyn:2018aa,Houzet:2019aa,Peralta-Gavensky:2019aa,Sakurai:2020aa,rui2021}.
However, the realization of topological materials turns out to be
challenging due to various reasons, such as a lack of tunability,
detrimental effects of impurities~\citep{Durand:2016aa,Jia:2011aa,Scanlon:2012aa,Takei:2013aa},
or quasiparticles in Majorana-based systems~\citep{Fu:2009aa,vanHeck2011,Rainis2012,Goldstein:2011aa,Budich:2012aa}.
A further challenge concerns the direct observability of the topological
invariant; often, the topological phase is only indirectly measured
through the density of states, e.g., via ARPES~\citep{Lv:2015aa}
or STM~\citep{Nadj-Perge:2014aa}.

\begin{figure}
	\centering
	\includegraphics[width=\linewidth]{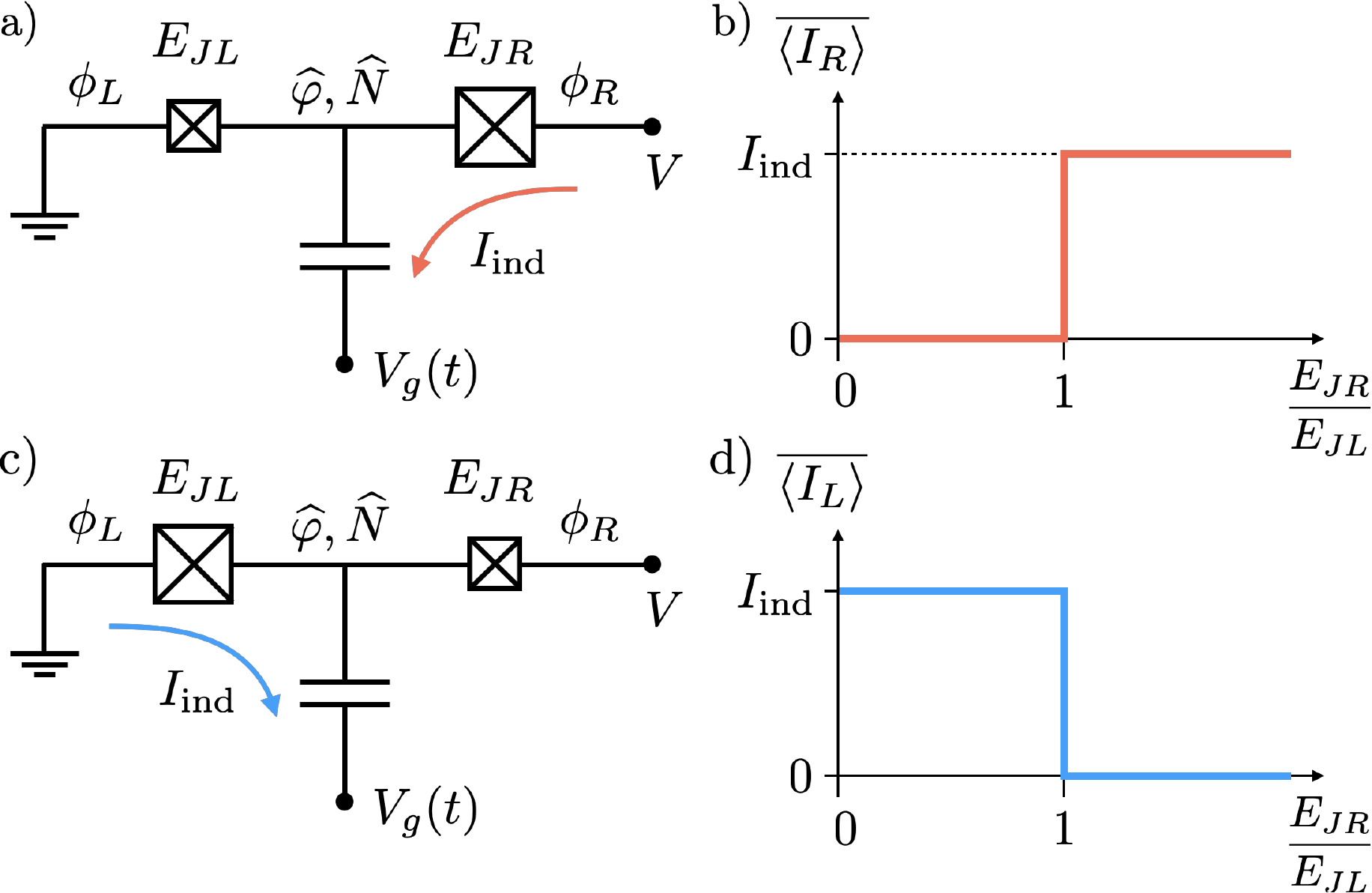}
	
	\caption{Circuit of the Cooper-pair transistor and quantized dc current responses.
		The two Josephson junctions are described by the energies $E_{JL}$ and $E_{JR}$. We apply a voltage $V$ between the left and right lead with superconducting phases $\phi_{L}$ and $\phi_{R}$, which drives the phase difference adiabatically according to the second Josephson relation. The charge and phase of the superconducting island are described by the conjugate variables $\widehat{N}$ and $\widehat{\varphi}$.
		The linearly time-dependent gate voltage $V_{g}\left(t\right)$ induces a dc current $I_{\text{ind}}\propto\dot{V}_{g}$ which flows a) entirely through the right junction if $E_{JR}>E_{JL}$ and c) entirely through the left junction if $E_{JR}<E_{JL}$.
		b) + d) The dc parts of the expectation values of the currents coming from the right and left lead, $\overline{\left\langle I_{R}\right\rangle }$ and $\overline{\left\langle I_{L}\right\rangle }$, respectively, are depicted under adiabatic conditions as a function of $ E_{JR} / E_{JL}$.
	}
	\label{fig:Circuit-and-currents}
\end{figure}

This is why alternatives are actively researched, where the topological
phase is encoded in other degrees of freedom. Circuit lattices~\citep{Ningyuan2015,Albert2015,Imhof:2018aa,Zhao_2018,Lu:2019aa,Rafi-Ul-Islam:2020aa,Wang:2020aa,Yu:2020aa,Lee:2018tn,Ezawa:2018uo,Li:2018uu}
may form metamaterials where topological numbers are defined through
the lattice degrees of freedom, which however requires the control
of a large number of circuits. Topological materials are also very
straightforwardly simulated when considering the space spanned by
the control parameters of superconducting qubits~\citep{Roushan:2014aa,Tan:2018aa,Tan:2019aa},
where it remains however unclear, how physics related to protected
edge states may be observed. Such limitations may be circumvented
by recently proposed topological transitions in multiterminal Josephson
junctions~\citep{Pankratova_2020_multitJJexp,Draelos2019multitJJexp,Heck2014TopolABSmultitJJ,Houzet:2019aa,weisbrich2021,Zazunov:2017aa,Xie:2019aa,Padurariu:2015aa,Peralta-Gavensky:2019aa,Sakurai:2020aa,Yang:2019aa,Yokoyama:2017aa}
which give rise to topological phases even when using only trivial
materials~\citep{Riwar2016,Repin:2019aa,Strambini:2016aa,Eriksson2017,Meyer:2017aa,Yokoyama2015TopolABSmultitJJ,Xie:2017aa,Xie:2018aa,Deb:2018aa,Klees:2020aa,Klees:2021aa}.
Here, Weyl points are found in the space of superconducting phase
differences acting as quasimomenta, and a Chern number can be directly
accessed through a quantized transconductance~\citep{Riwar2016}.
While topological transitions in transport degrees of freedom offer
a promising new approach, the proposal in Ref.~\citep{Riwar2016}
came with the experimental complication of needing small multiterminal
junctions containing only a few channels, which are at the same time
strongly tunnel-coupled. An important simplification was recently
proposed by means of Weyl points in standard SIS junction circuits~\citep{Fatemi:2021wm,Peyruchat:2021tb,Leone2008,Leone2008application,Leone_2013}.
However, these proposals require a control of the offset charges on
the order of a single electron charge $e$ which may be experimentally
challenging~\citep{Serniak2018,Serniak2019,Christensen2019}, unless
offset-charge feedback loops are employed~\citep{Pop2019}.

Here, we consider the Cooper-pair transistor, consisting of two tunnel
junctions with a superconducting island in between; see Fig.~\ref{fig:Circuit-and-currents}.
Although this circuit has been studied to a great extent~\citep{Cottet2002,Woerkom:2015aa,Korotkov1996,Eiles:1994aa,Giazotto:2015aa,Duty:2005aa,Aumentado:2004aa,Joyez:1994aa,Billangeon:2007aa,Proutski:2019aa,Arutyunov:2017aa,Naaman:2006aa,Rodrigues:2007aa,Chen:2014aa},
the Weyl points it exhibits in its band structure have, to the best
of our knowledge, not yet been discussed. Additionally to the phase
difference across the two junctions, we use the island offset charge
to define the Chern number which gives rise to a topological phase
transition when the asymmetry of the Josephson energies changes its
sign. This Chern number leads to a quantized dc current response into
a particular lead when driving both the offset charge and the phase
difference time dependently (see Fig.~\ref{fig:Circuit-and-currents}).

Remarkably, the quantization of the current response is \emph{insensitive
}to low-frequency offset-charge noise; in fact, it is actually beneficial
for the convergence of the response. Furthermore, we find that the
Chern number is insensitive to quasiparticle poisoning. Moreover,
the crossings in the higher bands occur via Weyl points with higher
topological charges. As a result, each band carries the same Chern
number such that it is not required to be in the ground state to observe
the effect. This is a surprising exception because usually in quantum
systems, the ground and excited states exhibit different topological
numbers, a fact which recently lead to the effort of generalizing
topological phase transitions to systems out of equilibrium~\citep{Kawabata:2019aa,McGinley:2019aa,McGinley:2018aa,Lieu:2020aa,Rudner:2009aa,Budich:2015aa,Bardyn:2018aa,Kunst:2018aa,Edvardsson:2019aa,Kastoryano:2019aa}.
Motivated by the above remarkable protection, we analyze the influence
from the environment more closely, by means of a generic master equation.
Based on this, we expect that the leading-order deviation of the quantized
current remains small. Finally, we discuss
an experimentally feasible protocol to measure the quantized current.
This protocol is to some extent reminiscent of Cooper-pair pumps~\citep{Niskanen2003,Vartiainen:2007aa,Gasparinetti2012,Pekola2013,Leone2008,Leone2008application,Leone_2013},
with the notable difference that previously studied pumps do suffer
from quasiparticle poisoning~\citep{Pekola2013}.

Compared to the above vast existing literature on topological quantum systems, the effect we study here has the following advantages. First, we do not need any topological materials nor a network of coupled circuits to find topological phase transitions -- a single circuit made of regular s-wave superconductors suffices. What is more, compared to similar recent proposals, this circuit has only two terminals and works already by means of standard SIS junctions. In fact, as we will show below, \textit{all} independent system parameters entering the Hamiltonian play a crucial role for the observed topological effect, leaving no ``spare'' degrees of freedom. It is in this sense that we refer to our system as a \emph{minimal} topological circuit.

This paper is structured as follows. In Sec.~\ref{sec:circuit-and-topology},
we review the Hamiltonian of the Cooper-pair transistor and discuss
its topological features. Afterwards, in Sec.~\ref{sec:Quantized-current-response},
we show how to access the Chern number by varying specific system
parameters in time. In Sec.~\ref{sec:open-system-description}, we
start with a short discussion of the robustness of the resulting quantized
dc current response with respect to the most common perturbations,
followed by the introduction of a generic open-system description
to discuss the leading-order perturbation to the otherwise protected
quantization of the current response. The concrete experimental realization
will be the topic of the final Sec.~\ref{sec:DC-current-measurement},
where we propose a practical measurement scheme.

\section{The circuit and its topology\label{sec:circuit-and-topology}}

\begin{figure}
	\centering
	\includegraphics[width=\linewidth]{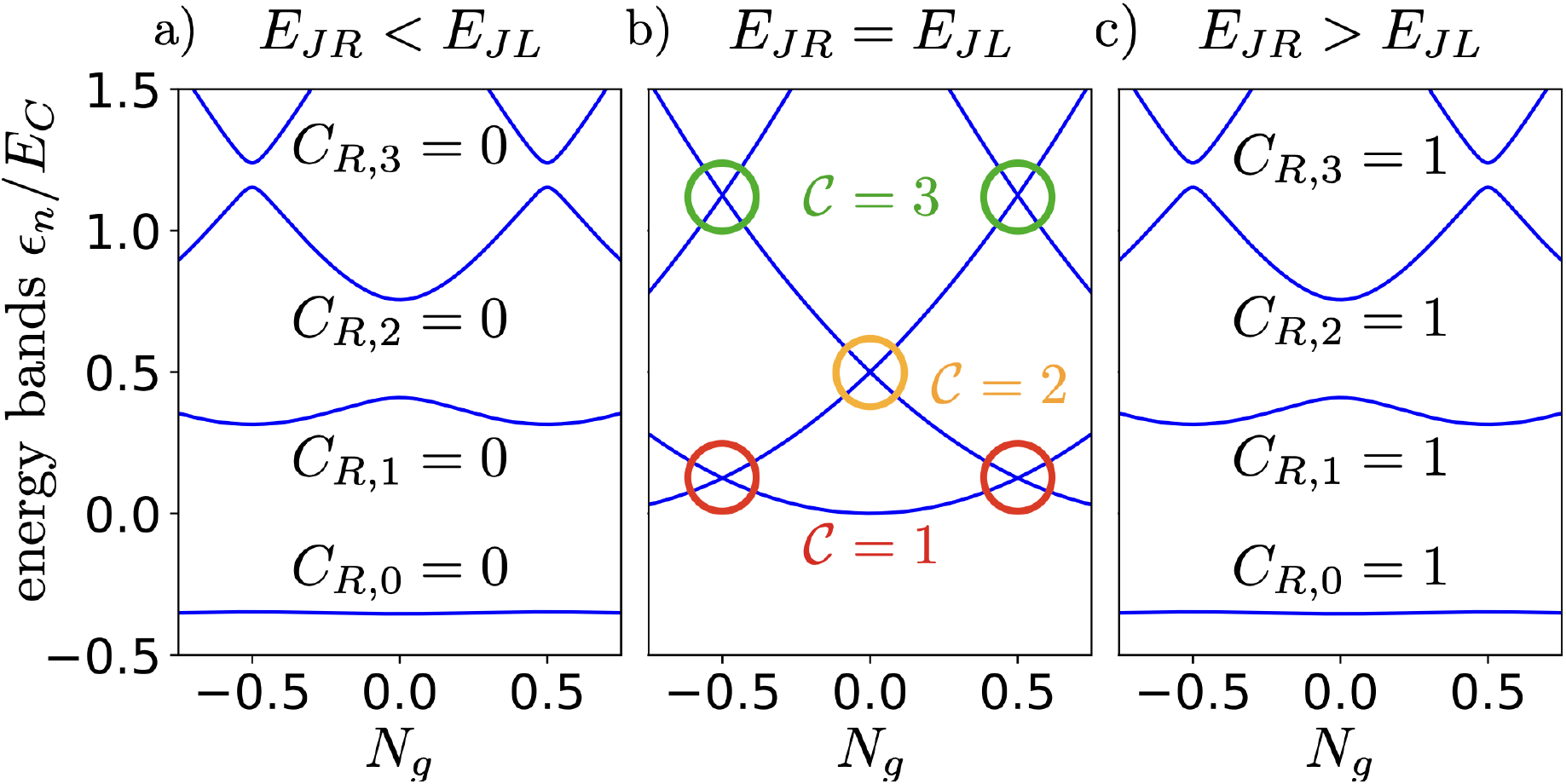}

	\caption{Energy spectrum and its topological properties.
		Displayed are the lowest four energy bands for $\phi = \phi_{R} - \phi_{L} = \pi$ as a function of the offset charge $N_{g}$ for three cases where the junction asymmetry changes from $E_{JR} < E_{JL}$ over $E_{JR} = E_{JL}$ to $E_{JR} > E_{JL}$.
		In the symmetric case, b), one can see the Weyl points as band crossings, each associated with a topological charge $\mathcal{C}$. In the asymmetric cases, a) and c), each band $n$ can be assigned a Chern number $C_{R,n}$, which is zero in the trivial phase and changes only when passing through a Weyl point. Due to the topological charges increasing by one with each higher Weyl point, the Chern numbers for the different bands all change by the same value.}
	\label{fig:Weyl_points_in_spectrum}
\end{figure}

We here investigate the topological properties of the Cooper-pair
transistor, consisting of two Josephson junctions connected in series
with energies $E_{JL}$ and $E_{JR}$ (see Fig.~\ref{fig:Circuit-and-currents}a)
forming a charge island, which is capacitively coupled to a gate voltage
$V_{g}$. The Hamiltonian is given by
\begin{align}
	\widehat{H}\left(N_{g},\phi_{L},\phi_{R}\right)= & \frac{E_{C}}{2}\left(\widehat{N}+N_{g}\right)^{2}-E_{JL}\cos\left(\widehat{\varphi}-\phi_{L}\right)\nonumber \\
 & -E_{JR}\cos\left(\widehat{\varphi}-\phi_{R}\right).\label{eq:Hamiltonian}
\end{align}
The first term describes the charging energy of the superconducting
island with the Cooper-pair number operator $\widehat{N}$ and $E_{C} = (2e)^{2} / (C_{L} + C_{R} + C_{g})$,
where $C_{L}$, $C_{R}$, and $C_{g}$ are the capacitances of the
two junctions and the gate capacitor, respectively. The gate voltage
induces the charge offset $N_{g} = C_{R} \dot{\phi}_{R} / (2e)^{2} + C_{L} \dot{\phi}_{L} / (2e)^{2} + C_{g} V_{g} / 2e$.
The phase operator $\widehat{\varphi}$ is canonically conjugate to the
number of Cooper pairs, such that $\bigl[\widehat{\varphi},\widehat{N}\bigr]=i$.
Due to charge quantization, we can write the Hamiltonian in the discrete
charge basis with $\widehat{N}=\sum_{N}N\left|N\right\rangle \left\langle N\right|$
and $e^{i\widehat{\varphi}}=\sum_{N}\left|N\right\rangle \left\langle N-1\right|$.
We denote the eigenenergies and eigenvectors of $\widehat{H}$ as
$\epsilon_{n}$ and $\left|n\right\rangle $, respectively, which
correspond to the standard Mathieu functions~\citep{Cottet2002,Koch:2007aa}.
The energy spectrum is shown in Fig.~\ref{fig:Weyl_points_in_spectrum}. 

While this system has already been extensively studied~\citep{Cottet2002,Woerkom:2015aa,Korotkov1996,Eiles:1994aa,Giazotto:2015aa,Duty:2005aa,Aumentado:2004aa,Joyez:1994aa,Billangeon:2007aa,Proutski:2019aa,Arutyunov:2017aa,Naaman:2006aa,Rodrigues:2007aa,Chen:2014aa},
it has to the best of our knowledge not yet been explicitly remarked
that it harbors nontrivial Chern numbers in the base space given by
the charge offset $N_{g}$ and the phase difference $\phi=\phi_{R}-\phi_{L}$,
\begin{equation}
	C_{\alpha,n}=\int_{0}^{1}\mathrm{d}N_{g}\int_{0}^{2\pi}\frac{\mathrm{d}\phi}{2\pi}\,B_{\alpha,n}\left(N_{g},\phi\right),\label{eq:Chern}
\end{equation}
with the Berry curvature $B_{\alpha,n} = -2 \operatorname{Im} \left\langle \partial_{\phi_{\alpha}} n \left| \partial_{N_{g}} n \right\rangle \right.$
($\alpha=\text{L},\text{R}$). In fact, the transistor thus simulates
a Chern insulator, where the parameter pair $\left(N_{g},\phi\right)$
acts as the Brillouin zone on a 2D torus~\footnote{We note that strictly speaking, the Hamiltonian is only periodic in
$\phi_{\alpha}$, but not in $N_{g}$, where an additional unitary
transformation appears, $\hat{H} \left(N_{g} + 1\right) = e^{-i\widehat{\varphi}} \hat{H} \left(N_{g}\right) e^{i\widehat{\varphi}}$.
However, it is easy to show that this transformation cancels in $B_{\alpha,n}$,
guaranteeing that the Chern number defined by Eq.~\eqref{eq:Chern}
remains quantized.}. Let us now explain the origin of the nontrivial Chern number in the remainder of this section.

First, we note that the Chern number defined in Eq.~\eqref{eq:Chern}, as well as the corresponding Berry curvature, carry the index \(\alpha=\text{L},\text{R}\), indicating that there are actually two distinct Chern numbers (Berry curvatures) for one and the same band. The reason for this is the following. The eigenenergies only depend on the total phase difference, $\epsilon_{n}\left(\phi\right)$,
such that $\partial_{\phi_{L}}\epsilon_{n} = -\partial_{\phi_{R}}\epsilon_{n}$.
This stems from the fact that $\phi_{\alpha}$ can be ``eliminated''
by the unitary transformation $\widehat{U}_{\alpha} \widehat{H} \widehat{U}_{\alpha}^{\dagger}$
with $\widehat{U}_{\alpha} = e^{i\phi_{\alpha} \widehat{N}}$ {[}e.g.,
$\widehat{U}_{L} \widehat{H}\left(N_{g}, \phi_{L}, \phi_{R}\right) \widehat{U}_{L}^{\dagger} = \widehat{H}\left(N_{g}, 0, \phi_{R} - \phi_{L}\right)${]}.
In other words, the eigenenergies are insensitive to a unitary
change of basis. The \emph{eigenvectors} on the other hand are not,
and therefore the two Berry curvatures $B_{L,n}$ and $B_{R,n}$ are
nontrivially related via
\begin{equation}
	B_{L,n} + B_{R,n} = -\partial_{N_{g}}\left\langle \widehat{N}\right\rangle _{n}\,. \label{eq:curvature_relation}
\end{equation}
This is a consequence of the current conservation law, $\sum_{\alpha} \partial_{\phi_{\alpha}} \widehat{H} = i\bigl[\widehat{H}, \widehat{N}\bigr]$,
which we will come back to in Sec.~\ref{sec:Quantized-current-response}.
Note that while $B_{\alpha,n}$ depends thus on $\alpha$, it depends
explicitly on only the phase difference, $B_{\alpha,n}\left(N_{g},\phi_{R}-\phi_{L}\right)$,
which is why it is sufficient to integrate over $\phi$ in Eq.~\eqref{eq:Chern}.

Based on Eq.~\eqref{eq:curvature_relation}, we can also relate the
Chern numbers for different $\alpha$,
\begin{equation}
	C_{L,n} + C_{R,n} = +1\,,\label{eq:Chern_relation}
\end{equation}
since $\int_{0}^{1} \mathrm{d}N_{g}\, \partial_{N_{g}} \bigl\langle \widehat{N} \bigr\rangle_{n} = -1$.
On a formal level, this difference between the Chern numbers for different
$\alpha$ is a simple consequence of the definition in Eq.~\eqref{eq:Chern}.
On a physical level, this difference will become meaningful in Sec.~\ref{sec:Quantized-current-response}, as the difference of measuring
the current from the left or from the right contact. In fact, the
physics will be engrained in the very structure of the Berry curvature:
a current measurement into contact $\alpha$ ($\partial_{\phi_{\alpha}}$)
as a response to a driving of $N_{g}$ ($\partial_{N_{g}}$).

The Chern numbers, as defined above, must be quantized, in accordance with standard literature~\citep{Thouless1982}. One way to explicitly calculate their actual values, would be by inserting the Mathieu functions into the Berry curvature and performing a numerical calculation of the values of the Chern numbers. However, we here resort to a simpler way which is inspired by Ref.~\citep{Riwar2016}.

Namely, the nonzero Chern numbers are a consequence of Weyl points (that is,
topologically protected band crossing points) appearing in the 3D
space given by $\left(N_{g}, \phi, E_{JR} / E_{JL}\right)$. These points
appear for symmetric junctions, $E_{JR} = E_{JL}$, at $\phi = \pi + 2 \pi m$,
where $m\in\mathbb{Z}$. Here, the Josephson energies for the left
and right junctions cancel in the Hamiltonian, such that only the
charging energy remains, $\widehat{H} = E_{C}\, \bigl(\widehat{N} + N_{g}\bigr)^{2} / 2$.
Therefore, the Weyl points simply represent the crossings of the shifted
parabolas for different charge states on the island. Indexing the
ground state as $n=0$ and the excited states with $n>0$ in ascending
order, one can state that the crossings between band $n$ and $n+1$,
for $n$ odd (even) occurs at $N_{g}$ being (half) integer; see Fig.~\ref{fig:Weyl_points_in_spectrum}b. Hence, when we tune from $E_{JR}<E_{JL}$
to $E_{JR}>E_{JL}$ (see Figs.~\ref{fig:Weyl_points_in_spectrum}a
and c), the Chern numbers for the different bands {[}Eq.~\eqref{eq:Chern}{]}
change according to the topological charge (or the chirality) $\mathcal{C}$
of the corresponding Weyl points (see Fig.~\ref{fig:Weyl_points_in_spectrum}b).

Importantly, while the Weyl points connecting the bands $n=0$ and
$n=1$ are regular Weyl points with topological charge $\mathcal{C}=+1$,
the Weyl points connecting arbitrary higher bands $n$ and $n+1$
have in fact a higher topological charge, $\mathcal{C}=+\left(n+1\right)$,
as indicated in Fig.~\ref{fig:Weyl_points_in_spectrum}b. We will
call such a point a multiple Weyl point, which can be considered as
a merger of $n+1$ regular Weyl points, each with charge $+1$, as
we explain in a moment. Since each band $n$ experiences a change
in its Chern number by subtracting the topological charge of the Weyl
point connecting to band $n-1$ from the topological charge of the
Weyl point connecting to $n+1$, the Chern numbers of \emph{all} the
bands are the same. For $C_{R,n}$, it follows that from a completely
trivial spectrum for $E_{JR}<E_{JL}$, where all $C_{R,n}=0$ (Fig.~\ref{fig:Weyl_points_in_spectrum}a), we go to a spectrum where all
bands have the same nonzero Chern number $C_{R,n}=+1$, for $E_{JR}>E_{JL}$
(Fig.~\ref{fig:Weyl_points_in_spectrum}c). Vice versa, for $C_{L,n}$,
we find $C_{L,n}=+1$ for $E_{JR}<E_{JL}$, and $C_{L,n}=0$ for $E_{JR}>E_{JL}$. Thus, it is only meaningful to consider a band as ``topologically trivial'' if we refer to a specific Chern number, \(C_{L, n}\) or \(C_{R, n}\), because one is zero and the other one is nonzero for every junction asymmetry, \(E_{JR} / E_{JL} \lessgtr 1\), independent of the band \(n\).
The Chern number being the same for all bands is a remarkable feature,
and renders the observation of the Chern number insensitive to whether
or not the system is in the ground state (e.g., when including finite
temperature); see also the discussion in Sec.~\ref{sec:open-system-description}.

Let us now discuss the physical origin of the multiple Weyl points.
We here provide an explicit discussion for the lowest two Weyl points,
which amount to the topological charges of $+1$ and $+2$. First,
regarding the regular Weyl point connecting the ground and first excited
state, the band crossing involves two charge states which differ by
only one Cooper pair, $\left|N-1\right\rangle $ and $\left|N\right\rangle $.
Tuning the parameters close to the crossing, $N_{g}=-N+1/2+\delta N_{g}$,
$E_{JR}=E_{JL}+\delta E_{J}$, and $\phi_{R}=\pi+\delta\phi$ while
at the same time $\phi_{L}=0$~\footnote{The physically relevant condition is of course $\phi_{R}-\phi_{L}=\pi+\delta\phi$.
The explicit choice made in the main text is merely to simplify the result, without loss of generality.}, we find the approximate Hamiltonian as derived in Appendix~\ref{sec:App-Single-Weyl-point}
\begin{equation}
	\widehat{H}_{1} = x\, \widehat{\sigma}_{x} + y\, \widehat{\sigma}_{y} + z\, \widehat{\sigma}_{z}\,,
	\label{eq:single_weyl}
\end{equation}
with $x=\delta E_{J}/2$, $y=E_{JL}\delta\phi/2$, and $z=E_{C}\delta N_{g}/2$,
and the Pauli matrices acting on the charge subspace \{$\left|N-1\right\rangle $,
$\left|N\right\rangle $\}, where $\widehat{\sigma}_{z}=\left|N\right\rangle \left\langle N\right|-\left|N-1\right\rangle \left\langle N-1\right|$.
This is the standard form of the Weyl Hamiltonian with the topological charge $\mathcal{C}=+1$.

As for the double Weyl point, with charge $+2$, we have to tune
to $N_{g}=-N+\delta N_{g}$, while the other small parameters ($\delta E_{J}$
and $\delta\phi$) are defined as above. Here, the relevant subspace
close to the crossing involves the charge states $\left|N-1\right\rangle $
and $\left|N+1\right\rangle $. Importantly, here it is impossible
to gap the two bands with the lowest-order Cooper-pair tunneling process,
because $\left\langle N-1\right|e^{\pm i\widehat{\varphi}}\left|N+1\right\rangle =0$.
Therefore, we need to go to higher-order processes involving the tunneling
of two Cooper-pairs via virtual charge states, which can be done by
means of a Schrieffer-Wolff transformation; see Appendix~\ref{sec:App-Double-Weyl-point}.
We find the Hamiltonian of the following form,
\begin{equation}
	\widehat{H}_{2} = z\, \widehat{\sigma}_{z} + 2 \left(x^{2} - y^{2}\right) \widehat{\sigma}_{x} + 4 x y\, \widehat{\sigma}_{y}\,,
	\label{eq:double_weyl}
\end{equation}
with $z=E_{C}\delta N_{g}$, $x=\delta E_{J}/2\sqrt{E_{C}}$, and $y=E_{JL}\delta\phi/2\sqrt{E_{C}}$, and where $\widehat{\sigma}_{z}=\left|N+1\right\rangle \left\langle N+1\right|-\left|N-1\right\rangle \left\langle N-1\right|$.
The fact that the Weyl point, here, has a topological charge of $+2$
can be shown in different ways. One could in principle define a Berry
curvature in the 3D space $\left(x,y,z\right)$ and compute explicitly
a closed surface integral enclosing the Weyl point. A more elegant
and instructive way is, however, to add a cotunneling term $E_{J}^{(2)}\cos\bigl(2\widehat{\varphi}\bigr)$
to the full Hamiltonian, Eq.~\eqref{eq:Hamiltonian}, which may originate
from higher-order tunneling processes in the SIS junction~\citep{Golubov2004cotunneling,Beenakker:1991vj,Smith:2020aa}.
Note that this cotunneling term is by no means relevant for our considerations. It
however serves as a neat mathematical trick to visualize the topological charge. Namely, this
term introduces a shift $c=E_{J}^{(2)}/4$ into $\widehat{H}_{2}$,
Eq.~\eqref{eq:double_weyl}, in the $\widehat{\sigma}_{x}$-term, $x^{2}-y^{2}\rightarrow x^{2}-y^{2}-c$.
As a consequence, the Weyl point at $\left(0,0,0\right)$ for $c=0$
splits into two regular Weyl points with topological charge $+1$
at $\left(\pm\sqrt{c},0,0\right)$ for finite $c$. Consequently,
without the splitting, it must have the topological charge $\mathcal{C}=+2$
and, thus, is a double Weyl point. This proof (not explicitly shown
here) can be extended to higher bands, where three or more regular
Weyl points are merged, giving rise to a triple or higher multiple
Weyl point, due to the gapping of the bands being third or higher
order in Cooper-pair tunneling processes, respectively.

To conclude, let us note one subtle difference between our notion of a Chern number compared to the more common definition within solid state theory, where the Chern number is usually connected to the single-particle band structure. Here, our bands are already in a many-body formulation, such that there is no need to associate occupation numbers to the individual bands.

\section{Quantized current response\label{sec:Quantized-current-response}}

We now show that the above discussed nonzero Chern numbers lead to
a directly measurable effect, which is a quantized, directed current
response, that is, a dc current flowing either precisely to the left,
or precisely to the right (depending on the junction asymmetry), as
depicted in Fig.~\ref{fig:Circuit-and-currents}. This effect emerges
when driving $N_{g}$ and $\phi$ time dependently. The driving of
the superconducting phase difference is accomplished by means of applying
a voltage,
\begin{equation}
	\dot{\phi}_{R} - \dot{\phi}_{L} = 2eV\,,
	\label{eq:constraint}
\end{equation}
whereas the gate-induced offset charge is linearly ramped up (or down)
with a constant ramping speed $\dot{N}_{g}$. Let us emphasize that there is an arbitrary
number of possibilities to satisfy Eq.~\eqref{eq:constraint}. Strictly
speaking, since these different choices are related through a time-dependent
unitary transformation $\widehat{U}$ (via \(\widehat{H} \rightarrow \widehat{U} \widehat{H} \widehat{U}^\dagger\)), they do not give rise to equivalent
Schr\"odinger equations, due to the additional term $-i\widehat{U} \partial_t \widehat{U}^\dagger$.
However, this leads merely to a shift in the initial condition
on $N_{g}(t)$, which is irrelevant for the here considered
dc current response (due to the time-averaging). Our subsequent results
are formulated independent of this choice.

Before proceeding, let us comment on one important point. Of course,
the ramping up of $N_{g}$ with a constant ramping speed can in reality
not be exerted for unlimited time, as the transistor would eventually
break. However, as we show in Sec.~\ref{sec:DC-current-measurement},
this is no actual limitation, as this problem can be easily circumvented
by choosing an appropriate driving protocol.

As we show now, the quantized current response resulting from the
time-dependent driving is in close analogy to the proposal in Ref.~\citep{Riwar2016}. In the four-terminal setup of Ref.~\citep{Riwar2016},
voltages were applied to two different contacts, resulting in a pure
$\phi$-driving and a resulting dc transconductance. Similar proposals
have very recently emerged in pure SIS junction circuits~\citep{Fatemi:2021wm,Peyruchat:2021tb}.
Here, on the other hand, we have a two-terminal device with only one
independent phase difference $\phi=\phi_{R}-\phi_{L}$ and driving
in the ``mixed'' parameter space $\left(N_{g},\phi\right)$.

To proceed, we consider the dynamics of the system for slow, adiabatic
driving. In this limit, the time-dependent Schr\"odinger equation $i\partial_{t}\left|\psi_{n}(t)\right\rangle =\widehat{H}(t)\left|\psi_{n}(t)\right\rangle $
has the solutions~\citep{Thouless1983}
\begin{equation}
	\left|\psi_{n}(t)\right\rangle = \mathrm{e}^{i\alpha_{n}(t)} \Biggl[\left|n(t)\right\rangle  + \sum_{m\neq n} \left|m(t)\right\rangle \frac{i\left\langle m(t)\right| \partial_{t} \left|n(t)\right\rangle }{\epsilon_{m}(t) - \epsilon_{n}(t)}\Biggr]\,,
	\label{eq:adiabatic_wave}
\end{equation}
under the assumption that, at the initial time $t_{0}$, the system
was in the eigenstate $\left|n(t_{0})\right\rangle $.
Here, $\widehat{H}(t) \left| n(t)\right\rangle =\epsilon_{n}(t) \left| n(t)\right\rangle $
denote the instantaneous eigenbasis at time $t$. The time evolution
gives rise to the (here irrelevant) dynamical phase $\alpha_{n}(t)=-i\int_{t_{0}}^{t}dt'\epsilon_{n}(t')$~\footnote{As in Ref.~\citep{Thouless1983} the instantaneous eigenbasis is chosen
such that the geometric phase contribution $\sim \left\langle n\right| \partial_{t} \left| n\right\rangle $ vanishes.}.
Note that for simplicity, we refrain from explicitly adding the
time argument $\left(t\right)$ from now on. This adiabatic approximation
is valid for~\citep{Mostafazadeh:1997aa} $\left|\left\langle m\right| \partial_{t} \left| n\right\rangle \right| \ll \left|\epsilon_{m} - \epsilon_{n}\right|$,
($n\neq m$), which in our case requires $\bigl|\dot{N}_{g}\bigr|, \left|V\right| \ll \inf |\epsilon_{m} - \epsilon_{n}| $. Also note that adiabaticity is a standard requirement (see also Refs.~\citep{Riwar2016,Eriksson2017,Fatemi:2021wm,Peyruchat:2021tb}) and that nonadiabatic effects like Landau-Zener transitions~\citep{Zener1932} are exponentially suppressed away from the degeneracy point.

As indicated above, we are interested in the expectation values of the currents
into the system from the left and right contacts in the presence of
the drive. The current operators are formally defined as
\begin{equation}
	\widehat{I}_{\alpha}=2e\,\partial_{\phi_{\alpha}}\widehat{H}\,.
	\label{eq:current_operator}
\end{equation}
The reason that we have to consider both the left and the right current
separately, is that the time-dependent driving of the gate charge
$N_{g}$ produces a finite dc displacement current, such that the
current expectation values do not simply add up to zero, as it would
be the case for the stationary system. Instead, we have to carefully
keep in mind the current conservation law
\begin{equation}
	\widehat{I}_{R} + \widehat{I}_{L} = 2e \dot{\widehat{N}} = 2e\, i \Bigl[\widehat{H}, \widehat{N}\Bigr]\,,\label{eq:current-conservation}
\end{equation}
where the right-hand side is nonzero, due to $\widehat{H}$ not commuting
with the island charge $\widehat{N}$. The expression $\dot{\widehat{N}}$
is, of course, to be understood in the Heisenberg picture.

We can now compute the expectation values of the currents, by inserting
the wave function given in Eq.~\eqref{eq:adiabatic_wave}. We find
\begin{equation}
	I_{\alpha,n} \equiv \left\langle \psi_{n}\right| \widehat{I}_{\alpha} \left|\psi_{n}\right\rangle
	= 2e \bigl[\partial_{\phi_{\alpha}} \epsilon_{n} + 2 \operatorname{Im}\left\langle \partial_{\phi_{\alpha}} n \left| \partial_{t} n \right\rangle \right.\bigr]\,,
	\label{eq:current_of_time}
\end{equation}
which we have written in the form $I_{\alpha, n} = I_{\alpha, n}^{(0)} + I_{\alpha, n}^{(1)}$. Here, $I_{\alpha, n}^{(0)} \sim \partial_{\phi_{\alpha}} \epsilon_{n}$ corresponds to a time-dependent version of the ordinary Josephson effect, while $I_{\alpha, n}^{(1)}$ is a correction term of first order in the driving parameters.

With the above result, we can now discuss the very important issue
of current conservation, to understand how the expectation values
of the left and right currents are related. The sum of the zero-order
contributions $\sum_{\alpha} I_{\alpha, n}^{(0)}$ has to
vanish due to the eigenenergies depending only on the total phase
difference (as discussed in Sec.~\ref{sec:circuit-and-topology}).
Importantly, however, the first-order term $I_{\alpha,n}^{(1)}$ gives
rise to a Berry curvature $B_{\alpha,n}$, as it appears in Eq.~\eqref{eq:Chern},
due to the driving of the offset charge $N_{g}$; and since the Berry
curvatures for different $\alpha$ being nontrivially related via
Eq.~\eqref{eq:curvature_relation}, the sum of the first-order contributions
is nonzero. This physical consequence is reflected in the current
conservation $\bigl\langle \widehat{I}_{R} + \widehat{I}_{L} \bigr\rangle_{\psi_{n}} = 2e\,\partial_{t} \bigl\langle \widehat{N} \bigr\rangle_{n}$,
in accordance with Eq.~\eqref{eq:current-conservation}.

In the same spirit as in Ref.~\citep{Riwar2016}, we now find that
when averaging the currents over long times $\overline{I}_{\alpha,n}=\lim_{\tau\rightarrow\infty}\int_{0}^{\tau}\frac{\mathrm{d}t}{\tau}I_{\alpha,n}$
(dc limit), the zeroth-order contributions $I_{\alpha,n}^{(0)}$
cancel and the first-order ones $I_{\alpha,n}^{(1)}$
average out to give the Chern numbers $C_{\alpha,n}$ from Eq.~\eqref{eq:Chern},
providing the result \(\overline{I}_{\alpha, n} = -2e \dot{N}_{g} C_{\alpha, n}\).
This is due to the currents being periodic in $N_{g}$ and $\phi$,
such that the long-time integral will eventually converge into an
area integral over the Brillouin zone of the $\left(N_{g},\phi\right)$-plane.
Importantly, the presence of the Weyl points and the resulting nontrivial
Chern numbers (as discussed above and shown in Fig.~\ref{fig:Weyl_points_in_spectrum})
lead to the quantized dc currents into the system
\begin{align}
	\overline{I}_{R, n} & =
	\begin{cases}
		0 & E_{JR} < E_{JL}\\
		-2e \dot{N}_{g} & E_{JR} > E_{JL}
	\end{cases}\,,
	\label{eq:IR_dc}\\
	\overline{I}_{L, n} & =
	\begin{cases}
		-2e \dot{N}_{g} & E_{JR} < E_{JL}\\
		0 & E_{JR} > E_{JL}
	\end{cases}\,.
	\label{eq:IL_dc}
\end{align}
This central result is also visualized in Fig.~\ref{fig:Circuit-and-currents}.
Namely, we find that the current injected into the system through
the ramping of the gate voltage, $2e\dot{N}_{g}$, is completely redirected
to the right (left) contact, if $E_{JR}$ is larger (smaller) than
$E_{JL}$ -- crucially, \emph{irrespective} of the precise ratio
between $E_{JR}$ and $E_{JL}$. As we see, the difference between
$C_{\alpha,n}$ for different $\alpha$ corresponds to current measurements
at different contacts $\alpha$, which are not equal due to the displacement
current.

We note that the plots in Fig.~\ref{fig:Circuit-and-currents} b) and d)
are idealized in the sense that it is purely hypothetical to stay adiabatic in the
vicinity of the symmetric point, \(E_{JR} = E_{JL}\), where we necessarily
come close to a degeneracy point. At such a point, Landau-Zener
transitions~\citep{Zener1932} cannot be avoided. These transitions, and
their importance for the observation of the Chern number have already been discussed
elsewhere, for SNS type junctions (see Ref.~\citep{Eriksson2017}), which is why we do not repeat a similar effort here. Overall, it can be expected that the step in Fig.~\ref{fig:Circuit-and-currents} b and d will be ``rounded'' due to Landau-Zener transitions, where the broadening decreases when reducing the driving frequency of \(N_g\) and \(\phi\). We also note
the importance of applying a voltage $V$ across the two contacts.
If we were to modulate $N_{g}$ only, the integral of the zeroth-order term
\(I_{\alpha,n}^{(0)}\sim\partial_{\phi_{\alpha}}\epsilon_{n}\)
would not in general vanish, nor would the integral of the Berry
curvature extend over the entire Brillouin zone of $\left(N_{g},\phi\right)$.
In the absence of any bias voltage, the total injected current through
the gate drive will be partitioned to the left and right contact with
proportions depending on many system parameters. The perfect directing
of the injected current to exclusively either the left or the right,
independent of the parameter details, is a pure topological effect,
requiring the combined modulation of the two parameters $N_{g}$ and
$\phi$.

Finally, let us get back to the titular notion of a \emph{minimal} topological circuit. In the Hamiltonian, there are overall the independent parameters \(E_C\), \(E_{JR} - E_{JL}\), \(\phi\), and \(N_g\)~\footnote{The energy \(E_{J} = E_{JL} + E_{JR}\) itself is not considered an independent parameter, as all energies can be rescaled with respect to a given reference energy. We can choose $E_J$ to be that reference.}. The last three out of these four parameters provide the base space in which the Weyl points are defined. Finally, $E_C$ guarantees that the energy spectrum is gapped, such that it is meaningful to consider discrete bands with individual Chern numbers. Thus, all involved parameters play an indispensable role for the observed topological effect.

\section{Stability with respect to external perturbations\label{sec:open-system-description}}

In analogy to Ref.~\citep{Riwar2016}, the convergence of the dc current
to the values in Eqs.~\eqref{eq:IR_dc} and \eqref{eq:IL_dc} requires
the driving frequencies $\dot{N}_{g}$ and $2eV$ either to be incommensurable
or to come with a sufficient amount of low-frequency noise to make
sure that the entire Brillouin zone is covered for sufficiently long
times $\tau$. Surprisingly, this means that, here, low-frequency
offset-charge noise actually \emph{helps} for the observation of the
Chern number, instead of perturbing it. This is a considerable advantage
with respect to other recent proposals~\citep{Fatemi:2021wm,Peyruchat:2021tb}
which rely on a control of the offset charge on the order of an elementary
charge $e$ during the entire integration time $\tau$, which seems
challenging given the experimental evidence for offset-charge noise~\citep{Serniak2018,Serniak2019,Christensen2019}.

A further important point concerns quasiparticle poisoning. Quasiparticles
appear to be much more numerous than what should be expected in equilibrium~\citep{Martinis2009,Riste2013} and induce stochastic switches between
states of different parity. They are harmful for a large number of
quantum devices such as Cooper-pair boxes~\citep{Lutchyn2005,Shaw2008},
transmon and fluxonium qubits~\citep{Catelani2011}, Flux-qubits~\citep{Leppakangas2012},
Majorana-based qubits~\citep{Fu:2009aa,vanHeck2011,Rainis2012,Goldstein:2011aa,Budich:2012aa},
or Cooper-pair sluices~\citep{Pekola2013}. Also, the observation
of transport Chern numbers defined purely in $\phi$-space~\citep{Riwar2016,Fatemi:2021wm,Peyruchat:2021tb}
is hampered by quasiparticle-induced parity flips, since the topological
numbers differ in different parity sectors. The Chern number we consider
here, however, is the same, independent of the parity. Namely, in
our formalism, parity flips can be accounted for by simple shifts
of $N_{g}$ by half an integer, $N_{g}\rightarrow N_{g}\pm1/2$. Due
to the periodicity of the Berry curvature in $N_{g}$-space, it follows
that $\int_{0}^{1}dN_{g}B_{\alpha,n}\left(N_{g}\pm1/2,\phi\right)=\int_{0}^{1}dN_{g}B_{\alpha,n}\left(N_{g},\phi\right)$,
demonstrating the insensitivity of the Chern number, Eq.~\eqref{eq:Chern},
on fermion parity.

Moreover, as already indicated above, the Chern numbers for all the
bands $n$ are the same, due to the higher Weyl points having higher
topological charges, such that the dc current does not depend on $n$,
see Eqs.~\eqref{eq:IR_dc} and \eqref{eq:IL_dc}. Therefore, the here
predicted effect is likewise not sensitive to finite temperature occupations
of higher energy bands, which is a further advantage over the proposals
in Refs.~\citep{Riwar2016,Fatemi:2021wm,Peyruchat:2021tb}.

Encouraged by these striking facts, we now consider the effects of
the environment in more detail. As we will show, stochastic transitions
induced by the environment will introduce what we expect to be a small
leading-order correction in the current response. Such stochastic
transitions may be of various origins. In particular, the aforementioned
beneficial effects of fluctuations in $N_{g}$ or $\phi_{\alpha}$
are restricted to low-frequency fluctuations, that is, in a frequency
regime where the noise can be considered adiabatic, such that it does
not give rise to stochastic transitions between different energy bands.
Even though the finite-frequency power spectrum of the noise can be
expected to be low, such transitions may still occur in reality and
have to be taken into account. Likewise, quasiparticles may induce
stochastic transitions within the time required to average the current
signal.

Therefore, a detailed open quantum system description would have to
encompass a large number of models to account for all perturbations.
For instance, phase fluctuations are standardly described by an external
impedance, which can be modeled by an ensemble of
$LC$-resonators~\citep{Ingold:1992aa}. Charge fluctuations are usually modeled via
so-called two-level fluctuators~\citep{Dutta1981,Kenyon2000,Muller2019}.
However, such models have recently been put into question for 2D transmons,
where a deviation from the typical $1/f$-noise spectrum has been
observed~\citep{Christensen2019}. In order to avoid any dependence
on such details, we here take into account the open-system dynamics
as generally as possible, by means of a quantum master equation for
the density matrix of the system $\widehat{\rho}$,
\begin{equation}
	\partial_{t}\widehat{\rho}=-i\left[\widehat{H}(t),\widehat{\rho}\right]+\mathbf{W}(t)\,\widehat{\rho}\,.\label{eq:master-equation}
\end{equation}
The effect of the environment is described by the time-local kernel
$\mathbf{W}$, which may in general be time-dependent (since the system
is driven time dependently). In order to guarantee positivity of $\widehat{\rho}$
for all times $t$, we assume that $\mathbf{W}$ can be cast into
a Lindblad form (whose specific form however is, in a first approach,
irrelevant). Apart from that, we assume only that in the absence of
the time-dependent driving, the system will, up to small
corrections, end up in the ground state as the stationary solution,
$\rho^{\text{st}}\approx \widehat{\rho}_{0} = \left|0\right\rangle \left\langle 0\right|$.
This is equivalent to assuming that the environment has a small temperature
with respect to the band gaps, $k_{\text{B}}T \ll \inf |\epsilon_{n}-\epsilon_{0}|$.
Thus, we can understand $\mathbf{W}$ as a generic cooling mechanism. Note importantly that we do not assume the usually standard rotating-wave approximation (RWA) and instead keep processes in $\mathbf{W}$ which couple diagonal and off-diagonal contributions to the density matrix (that is, diagonal and off-diagonal with respect to the instantaneous eigenbasis of $\widehat{H}$). Strikingly, we find in the absence of the RWA a leading-order contribution which would otherwise be neglected and which is, at least nominally, of the same order as the Thouless result. Nonetheless, this correction can be shown to be small, due to other small factors; see below.

Let us now again focus on slow driving. For this purpose it is useful to cast Eq.~\eqref{eq:master-equation} into the aforementioned instantaneous eigenbasis of $\widehat{H}$. We thus get
\begin{equation}
	\partial_{t}\widehat{\rho}=-i\left(\mathbf{L}_0 + \delta\mathbf{L}\right) \widehat{\rho} + \mathbf{W}(t)\,\widehat{\rho}\,.
	\label{eq:master-eq-instantaneous}
\end{equation}
Importantly, all the objects appearing in Eq.~\eqref{eq:master-eq-instantaneous} differ from the ones in Eq.~\eqref{eq:master-equation} by the time-dependent unitary transformation $\widehat{U}(t)$, which changes the basis to the instantaneous eigenbasis. For example, for the density matrix, this corresponds to a mapping $ \widehat{\rho} \rightarrow \widehat{U} \widehat{\rho}\, \widehat{U}^\dagger $.
Therefore, one should strictly speaking use different symbols for Eqs.~\eqref{eq:master-equation} and \eqref{eq:master-eq-instantaneous} from which we will refrain for notational simplicity. As a consequence of the unitary transformation, the closed system dynamics receives an extra term to the ordinary $ \mathbf{L}_{0} \bullet = \bigl[\widehat{H}, \bullet\bigr] $, denoted as $ \delta\mathbf{L} \bullet = \bigl[-i\widehat{U} \partial_{t} \widehat{U}^\dagger, \bullet\bigr] $~\footnote{This notation of the ``$\bullet$'' is used to define superoperators mapping from one operator to another one, where the ``$\bullet$'' is to be replaced by the operator on which the superoperator acts.}.

For a consistent slow-driving approximation we have to consider the relationship between the three time scales $\left\Vert \mathbf{W}\right\Vert^{-1} $, $\left\Vert \delta\mathbf{L}\right\Vert^{-1} $, and $\left\Vert \mathbf{L}_0\right\Vert^{-1} $ (where $\left\Vert \mathbf{A}\right\Vert $ is a suitably chosen norm to capture the magnitude of $\mathbf{A}$). The regular closed-system dynamics scales with the instantaneous energy gaps $\left\Vert \mathbf{L}_{0}\right\Vert \sim\left|\epsilon_{m}-\epsilon_{n}\right|$ and the additional contribution due to the driving scales as $\left\Vert \delta\mathbf{L}\right\Vert \sim\left\langle m\right|\partial_{t}\left|n\right\rangle$ (both with $m\neq n$). Now, the regime of interest is $ \left\Vert \mathbf{L}_{0} \right\Vert^{-1} < \left\Vert \mathbf{W} \right\Vert^{-1} < \left\Vert \delta\mathbf{L} \right\Vert^{-1} $. Thus, we explicitly expand the density matrix in orders of $\left\Vert \mathbf{W}\right\Vert$ and $\left\Vert \delta\mathbf{L}\right\Vert$,
\begin{equation}
	\widehat{\rho} = \sum_{\nu, \mu} \widehat{\rho}^{(\nu, \mu)}\,, \label{eq:expanded_density_matrix}
\end{equation}
where $\widehat{\rho}^{(\nu, \mu)}$ scales as $ \left\Vert \delta\mathbf{L}\right\Vert^{\nu} \left\Vert \mathbf{W}\right\Vert^{\mu} $, and plug the result into the expectation value $I_\alpha = \operatorname{tr} \bigl[\widehat{I}_\alpha \widehat{\rho}\bigr]$. The derivation of these results can be found in Appendix~\ref{sec:App-Open-system-corrections}. Note that in general, there may be specific types of system-reservoir interactions providing an additional contribution to the current, which can generically be taken into account by a ``current kernel'' $\mathbf{W}_{I}$~\citep{Konig:1999td}, such that $I_\alpha = \operatorname{tr} \bigl[\widehat{I}_\alpha \widehat{\rho}\bigr] + \operatorname{tr} \bigl[\mathbf{W}_{I} \widehat{\rho}\bigr]$. For simplicity, we assume that the system-reservoir interaction is current conserving such that there is no such $ \mathbf{W}_{I} $ term. Since $ \left\Vert \mathbf{W}\right\Vert > \left\Vert\delta\mathbf{L}\right\Vert $, the corrections of $\widehat{\rho}$ due to the open-system dynamics are in principle dominant. However, in the absence of an explicit (time-dependent) driving and the environment at equilibrium, the system cannot generate a finite dc current, such that the dc contributions of all $I_\alpha^{(0, \mu)} = \operatorname{tr} \bigl[\widehat{I}_\alpha \widehat{\rho}^{(0, \mu)}\bigr]$ must be zero~\citep{footnote02}. Therefore, we need to consider at least first order in the driving parameter $\left\Vert\delta\mathbf{L}\right\Vert $. As we will show in the following, the leading-order terms of the dc current will be
\begin{equation}
	\overline{I}_\alpha = \overline{I}_{\alpha}^{(1,0)} = \overline{I}_{\alpha, o}^{(1,0)} + \overline{I}_{\alpha, d}^{(1,0)}\,,
	\label{eq:dc-current_open}
\end{equation}
the computation of which is detailed in Appendix~\ref{sec:App-Open-system-corrections}. Nominally, both of these contributions, \(\overline{I}_{\alpha, o}^{(1,0)}\) and \(\overline{I}_{\alpha, d}^{(1,0)}\), are of the same order \((1,0)\). However, they are of fundamentally different origin. As we will discuss now, the first term, $\overline{I}_{\alpha, o}^{(1,0)} = \overline{I}_{\alpha, 0}^{(1)}$, is simply the closed-system current proportional to the Chern number as computed previously in Sec.~\ref{sec:Quantized-current-response}. The second term, \( \overline{I}_{\alpha, d}^{(1,0)}\), will provide the leading-order correction to the above Chern number term,
and is due to a combination of driving and open-system dynamics (even though it is nominally of zero order in \(\Vert \mathbf{W} \Vert\)), as we will see in a moment. Moreover, its existence is based on the kernel creating transitions between diagonal and off-diagonal matrix elements, which is not the case when applying the RWA.

As already foreshadowed in Eq.~\eqref{eq:dc-current_open}, in order to discuss the current it is useful to decompose each order, \(I_\alpha ^{(\nu, \mu)} = \operatorname{tr} \bigl[\widehat{I}_\alpha \widehat{\rho}^{(\nu, \mu)}\bigr]\), as
\begin{equation}
	I_{\alpha}^{(\nu, \mu)} = I_{\alpha, d}^{(\nu, \mu)} + I_{\alpha, o}^{(\nu, \mu)}\,,
	\label{eq:current_open}
\end{equation}
with
\begin{equation}
	 I_{\alpha, d/o}^{(\nu, \mu)} = \operatorname{tr} \bigl[\widehat{I}_\alpha \mathbf{P}_{d/o}\, \widehat{\rho}^{(\nu, \mu)}\bigr]\,.
\end{equation}
Here, the projection superoperator $\mathbf{P}_{d}$ is defined to project onto the diagonal sector (in the eigenbasis of \(\widehat{H}\)), giving \(\langle n\vert (\mathbf{P}_d\, \widehat{\rho}\,) \vert m\rangle = \delta_{nm} \langle n\vert \widehat{\rho}\, \vert m\rangle\). Vice versa, $\mathbf{P}_{o}$ projects onto the off-diagonal subspace, such that \(\mathbf{P}_{d} + \mathbf{P}_{o} = \mathbf{I}\), where \(\mathbf{I}\) is the identity superoperator, leaving any input matrix unchanged.

Thus, the first term,
\begin{equation}
	I_{\alpha, d}^{(\nu, \mu)} = 2e \sum_{n} \rho_{nn}^{(\nu, \mu)}\, \partial_{\phi_\alpha} \epsilon_n\,,
	\label{eq:diagonal_order-by-order_current}
\end{equation}
is arising from the density-matrix contributions that are diagonal (analogously to the zero-order term in the closed system, \(I_{\alpha, 0}^{(0)} = \partial_{\phi_\alpha} \epsilon_n\)), while the second term is associated with the off-diagonal part of \(\widehat{\rho}\), originating in the driving and the interaction with the environment. It can be brought into the form
\begin{equation}
	I_{\alpha, o}^{(\nu, \mu)} = 2e \operatorname{tr} \left[ i \widehat{N}_{\alpha} \mathbf{L}_0 \widehat{\rho}^{(\nu, \mu)} \right].
	\label{eq:current_off-diag-correction}
\end{equation}
Here, we defined $\widehat{N}_{\alpha} \equiv -i\sum_{n, m} \left\langle n\right| \partial_{\phi_{\alpha}} \left|m\right\rangle \left|n\right\rangle \left\langle m\right|$. Evidently, $\widehat{N}_{\alpha}$ can be formally related to the Cooper-pair number operator of contact $\alpha$. However, note that care has to be taken with this interpretation. The contacts in the here considered model are macroscopically large and their actual charge operators do not have a well-defined expectation value, whereas $\widehat{N}_{\alpha}$ is always well-behaved. To avoid such unnecessary complications, we simply refer to it in the way it is defined: as the operator $-i\partial_{\phi_{\alpha}}$ expressed in the eigenbasis of $\widehat{H}$.

We now have to find the contributions to the density matrix by expanding it according to Eq.~\eqref{eq:expanded_density_matrix} and solving the Lindblad equation, Eq.~\eqref{eq:master-eq-instantaneous}, for leading orders in \((\nu, \mu)\). The details of this calculation are shown in Appendix~\ref{sec:App-Open-system-corrections}. Let us first consider the contribution \(I_{\alpha, o}\) as defined in Eq.~\eqref{eq:current_off-diag-correction}. When plugging in the solution \(\widehat{\rho}^{(1,0)}\), we find the term
\begin{equation}
	I_{\alpha, o}^{(1, 0)} = -2e\operatorname{tr}\left[i \widehat{N}_{\alpha} \delta\mathbf{L} \widehat{\rho}_{0}\right] = I_{\alpha, 0}^{(1)}\,,
	\label{eq:current_1,0-contribution_Berry-curvature-term}
\end{equation}
giving rise to exactly the same Berry curvature term as in the closed system; see Eq.~\eqref{eq:current_of_time}. For \(I_{\alpha, o}\) one could now in principle compute higher-order terms, such as \(\widehat{\rho}^{(1,1)}\), to obtain corrections due to the open-system dynamics. However, as already indicated in Eq.~(17), when including the impact of the environment, it turns out that there will actually be a correction nominally of zero order in \(\Vert \mathbf{W} \Vert\) in the current contribution \(I_{\alpha, d}\).

Let us here discuss the origin of this surprising new term. When treating the closed system in Sec.~\ref{sec:Quantized-current-response}, we have given the solution of the diagonal part of the density matrix as \(\mathbf{P}_d\, \widehat{\rho} = \vert 0\rangle \langle 0\vert\). However, strictly speaking, this is not the most general solution for the closed system. As a matter of fact, in the absence of any relaxation mechanisms, the most general solution for the diagonal part can be any mixed state \(\mathbf{P}_d\, \widehat{\rho} = \sum_n \rho_{nn} \vert n\rangle \langle n\vert\) with arbitrary \(\rho_{nn}\) \emph{provided} that \(\rho_{nn}\) be constant in time. If all \(\rho_{nn}\) are indeed constant in time, then Eq.~\eqref{eq:diagonal_order-by-order_current} will average to zero when computing the dc component. In the open system, the kernel \(\mathbf{W}\) accomplishes two things. On the one hand, it fixes the solution for the diagonal part of the density matrix, that is the \(\rho_{nn}\) are now unique. On the other hand, these \(\rho_{nn}\) now depend in general on time. Let us emphasize that this nontrivial behavior of the diagonal part is present only when including processes that couple the diagonal and the off-diagonal sectors in \(\mathbf{W}\) -- in other words, it is a consequence of going beyond the RWA. Therefore, we find
\begin{equation}
	I_{\alpha, d}^{(1, 0)} = 2e \sum_{n} \rho_{nn}^{(1, 0)}\, \partial_{\phi_\alpha} \epsilon_n\,,
	\label{eq:dominant-open-system-correction}
\end{equation}
with
\begin{multline}
	\rho^{(1, 0)}_{nn} = \left[\mathbf{\widetilde{W}}_{dd}^{-1} \mathbf{P}_{d} \left(\delta \mathbf{L} \frac{1}{\mathbf{L}_{0}} \mathbf{W} + \mathbf{W} \frac{1}{\mathbf{L}_{0}} \delta \mathbf{L}\right.\right.\\
	\left.\left.+ i\partial_{t} \Bigl[\widetilde{\mathbf{W}}_{dd}^{-1} \mathbf{P}_{d} \mathbf{W} \frac{1}{\mathbf{L}_{0}} \mathbf{W}\Bigr] \right) \widehat{\rho}_{0}\right]_{nn}.
	\label{eq:diag-1-0-density-matrix}
\end{multline}
Since $\partial_t \rho_{nn}^{(1,0)} \neq 0$ in general, the dc contribution of this current will be nonzero, marking it as the leading-order correction term.
Note that the inverse $1 / \mathbf{L}_{0}$ is defined only in the off-diagonal sector. However, the objects $ \mathbf{W} \widehat{\rho}_{0} $ and $ \delta\mathbf{L} \widehat{\rho}_{0} $, on which this inverse acts, are both purely off-diagonal, which is also explained in the Appendix~\ref{sec:App-Open-system-corrections}. Similarly, \(\mathbf{W}_{dd} \equiv \mathbf{P}_{d} \mathbf{W} \mathbf{P}_{d}\) has a zero eigenvalue, corresponding to a degenerate eigenspace including the ground state, \(\widehat{\rho}_0 = |0\rangle \langle 0|\), and any purely off-diagonal matrix.
Therefore, its inverse is ill-defined. However, we can construct the inverse \(\mathbf{\widetilde{W}}_{dd}^{-1}\) such that \(\mathbf{\widetilde{W}}_{dd}^{-1} \mathbf{W}_{dd}\, \bullet = \mathbf{P}_{d} \bullet - \widehat{\rho}_0 \operatorname{tr}[\bullet]\).
For more details see Appendix~\ref{sec:App-Open-system-corrections}.

While Eqs.~\eqref{eq:dominant-open-system-correction} and \eqref{eq:diag-1-0-density-matrix} are general under the assumptions made, for the sake of concreteness, we now compute them numerically considering background charge fluctuations~\citep{Dutta1981,Kenyon2000,Muller2019}, $N_g \rightarrow N_g + \delta\widehat{N}_g$. Here, $\delta\widehat{N}_g$ is an operator, whose dynamics is governed by an environment Hamiltonian, \(\widehat{H}_{\text{env}}\). The circuit Hamiltonian can now be approximated as follows, \(\widehat{H}(N_g) \rightarrow \widehat{H}(N_g + \delta \widehat{N}_g) \approx \widehat{H}(N_g) + \widehat{V}\), where we find the interaction with the environment \(\widehat{V} = \gamma E_C \widehat{N} \otimes \delta \widehat{N}_g\). The total system plus environment is thus described by the Hamiltonian $\widehat{H}(N_g) + \widehat{V} + \widehat{H}_{\text{env}}$.
Let us now focus on the regime where \(E_J(\phi) \gg E_C\) with the Josephson energy \(E_J(\phi) = \sqrt{E_{JR}^2 + E_{JL}^2 + 2 E_{JR} E_{JL} \cos(\phi)}\) (which we refer to as transmon limit).
This regime is of particular interest for the here considered topological effect, because the energy bands are flatter and therefore transitions between bands (via Landau-Zener) are suppressed. Note that we aim to remain in this transmon regime for all \(\phi\), which implies in addition, that the two junctions should be sufficiently asymmetric w.r.t.~\(E_C\), \(\left\vert E_{JR} - E_{JL} \right\vert \gg E_{C}\). Additionally, we assume \(\xi \ll 1\), with \(\xi \equiv \inf \{E_{JR} / E_{JL}, E_{JL} / E_{JR}\}\), to simplify the numerical calculations.
In this transmon limit, we can describe our system as a damped harmonic oscillator (HO) with an energy spacing of \(\omega_0 = \sqrt{E_C E_J(\phi)}\). The respective kernel is a well-known result from standard literature of open quantum systems~\citep{Maniscalco:2004ux}, which can be computed by tracing out the environment degrees of freedom, resulting in
\begin{equation}
	\mathbf{W} = - \Delta\, \mathbf{L}_{P}^{2} - \Pi\, \mathbf{L}_{P} \mathbf{L}_{X} + \frac{i}{2} r\, \mathbf{L}_{P^2} + i \zeta\, \mathbf{L}_{P} \overline{\mathbf{L}}_{X}\,,
\end{equation}
with \(\mathbf{L}_{A} \bullet = \bigl[\widehat{A}, \bullet \bigr]\), \(\widehat{X} = \sqrt[4]{E_J / E_C}\, \bigl[ \widehat{\varphi} - \phi_{L} - \delta(\phi) \bigr]\), \(\delta(\phi) = \arctan \bigl[\sin \phi / (E_{JL} / E_{JR} + \cos \phi)\bigr]\), and \(\widehat{P} = \sqrt[4]{E_C / E_J}\, \bigl(\widehat{N} + N_g\bigr)\). The four (real) correlation functions are
\begin{align}
	\Delta - i\Pi &= \int_{-\infty}^{0} \mathrm{d}t_1\, \kappa(t_1)\, \mathrm{e}^{i\omega_0 t_1}\,,\\
	-\frac{r}{2} + i\zeta &= \int_{-\infty}^{0} \mathrm{d}t_1\, \mu(t_1)\, \mathrm{e}^{i\omega_0 t_1}\,,
\end{align}
where
\begin{align}
	\kappa(t_1) &= \frac{\lambda^2}{2} \left\langle \left\{ \delta \widehat{N}_{g}(t_1), \delta \widehat{N}_{g}(0)\right\} \right\rangle ,\\
	\mu(t_1) &= i \frac{\lambda^2}{2} \left\langle \left[ \delta \widehat{N}_{g}(t_1), \delta \widehat{N}_{g}(0)\right] \right\rangle ,
\end{align}
with \(\lambda = \gamma \sqrt{E_C \omega_0}\). We observe that these correlation functions depend only on the transmon frequency \(\omega_0\) (and \(E_C\)), which in turn does not depend on \(N_g\) and only very weakly on \(\phi\). Therefore, this already indicates that the last term of Eq.~\eqref{eq:diag-1-0-density-matrix}, i.e., the term with the time derivative \(\partial_t\), is not the dominant term.

\begin{figure}
	\centering
	\includegraphics[width=\linewidth]{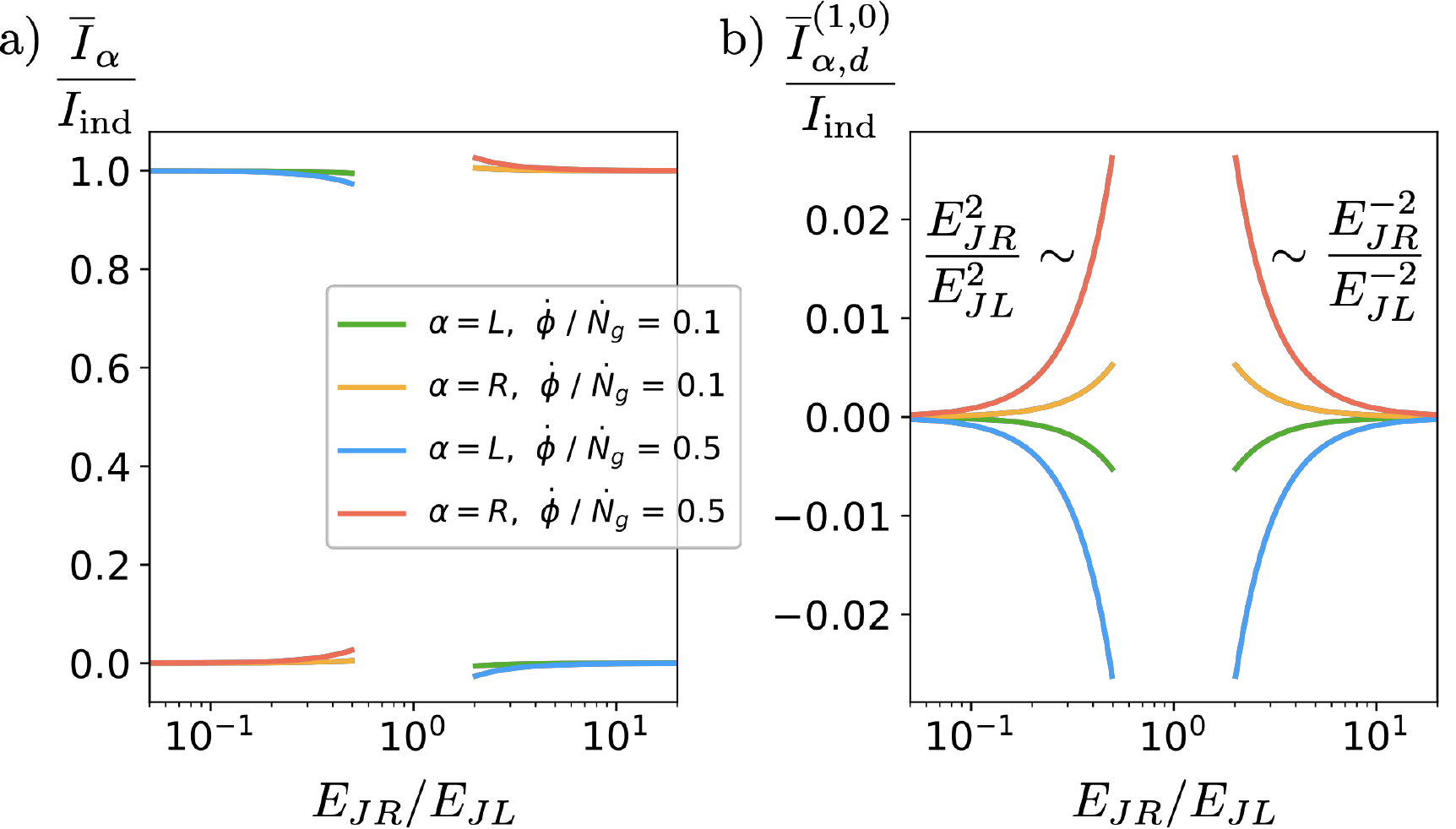}
	\caption[Open-system dc current and deviation from closed-system quantization]{Open-system dc currents.
		Depicted are a) the full dc current response in the open system, \(\overline{I}_{\alpha}\), and b) the leading-order correction of the response due to the open-system description, \(\overline{I}_{\alpha, d}^{(1, 0)}\), both relative to the current \(I_{\text{ind}} = -2e \dot{N}_g\), induced by the ramping of the offset charge. Those numerical results are obtained by assuming a large junction asymmetry, \(E_{JR} / E_{JL} \gg 1\) or \(E_{JR} / E_{JL} \ll 1\). In the intermediate regime, the perturbation cannot be expected to remain small (or even finite) due to the energy gap decreasing to zero when approaching the point degeneracy. For a large asymmetry, the deviation is quadratically suppressed.
		We chose the parameters \(E_{JR} + E_{JL} = 50 E_C\), \(\omega_c = 100 E_C\), and either \(\dot{\phi} = 0.1 \dot{N}_g\) or \(\dot{\phi} = 0.5 \dot{N}_g\) to demonstrate that the current correction scales with \(\dot{\phi} = 2eV\), and not with \(\dot{N}_g\) as for the closed-system current.}
	\label{fig:open-system_current}
\end{figure}

Due to microreversibility, the ratio of excitation and relaxation rates of two neighboring states is a Boltzmann factor \(\Gamma_{n\rightarrow n+1} / \Gamma_{n+1\rightarrow n} = (\Delta - \zeta) / (\Delta + \zeta) = \mathrm{e}^{-\beta\omega_{0}}\), with \(\Gamma_{n\rightarrow n^{\prime}} \equiv \langle n^{\prime}\vert \left( \mathbf{W} \vert n\rangle \langle n\vert \right) \vert n^{\prime}\rangle\). And since we focus on low temperature (compared to \(\omega_0\)), we find that \(\Delta \approx \zeta\). Thus, we have three independent correlation functions determining the kernel, which, for realistic predictions, would have to be identified experimentally along similar lines as Ref.~\citep{Christensen2019}. Here, for demonstration purposes (to show that the above discussed correction does not vanish in general), we assume the reservoir to be an ohmic bath with a cutoff frequency \(\omega_c \gg \omega_0\). Herewith, one finds the following scalings of the correlation functions~\citep{Maniscalco:2004ux}
\begin{align}
	\Delta, \zeta &\sim \gamma^{2} E_{C} \frac{\omega_{0}^{2}}{\omega_{c}^{2}}\,,\\
	\Pi &\sim \gamma^{2} E_{C} \frac{\omega_{0}^{2}} {\omega_{c}^{2}} \ln\biggl(\frac{\omega_{0}}{\omega_{c}}\biggr)\,,\\
	r &\sim \gamma^{2} E_{C} \frac{\omega_{0}}{\omega_{c}}\,.
\end{align}
Assuming a junction asymmetry described by \(\xi \ll 1\) allows us now to expand the current correction from Eq.~\eqref{eq:dominant-open-system-correction} w.r.t.~\(\xi\). Since all \(\phi\)-dependencies enter via a dependence on \(E_J(\phi)\) or \(\partial_{\phi} \delta(\phi) = E_{JR} \left(E_{JR} + E_{JL} \cos\phi \right) / E_J^2(\phi)\), we find that the zeroth order (in \(\xi\)) is always a constant term (in \(\phi\)) while the first order is associated with a harmonic dependence. Thus, odd orders of the current vanish when averaging over \(\phi\) (corresponding to a time integral), and since \(\partial_{\phi_\alpha} \epsilon_n\) has a vanishing zeroth order, the correction has to be at least of quadratic order in the asymmetry \(\xi\).
This can be confirmed numerically: see Fig.~\ref{fig:open-system_current}, where we explicitly expand Eq.~\eqref{eq:dominant-open-system-correction} up to second order in \(\xi\), and subsequently perform an integration over \(\phi\) to obtain the dc component of \(I_{\alpha, d}^{(1,0)}\).

We conclude that while the dc part of the open system correction to the current does not vanish, and gives rise to a deviation from the otherwise perfect current quantization, our analysis offers very concrete indications how to minimize its influence. Namely, by means of our general discussion above, we identify an important difference in the scaling behavior between the topological part of the current response, and the open system correction: while the
former appears with a prefactor proportional to $\dot{N}_{g}$
{[}see Eqs.~\eqref{eq:IR_dc} and \eqref{eq:IL_dc}{]},
the correction in Eq.~\eqref{eq:dominant-open-system-correction} turns out to scale with $\dot{\phi} = 2eV$, such that for this deviation to be small, $V$ should not be chosen too large with respect to the ramping of $N_{g}$.
To conclude, we find that the correction is small as long as \(\xi^2\, V /\dot{N}_{g} < \left\Vert \mathbf{L}_{0} \right\Vert / \left\Vert \mathbf{W} \right\Vert\). Therefore, while the open system correction is not exponentially suppressed (which seems to be a generic feature of open systems; see, e.g., a recent discussion for topological insulators~\citep{McGinley2020}), our above calculation provides very clear and stringent strategies to mitigate it, by choosing the driving and other parameters accordingly. In particular, as we explicitly show in Fig.~\ref{fig:open-system_current}, the overall open-system correction can be kept very small, due to \(\xi \ll 1\) and \(\dot{\phi} < \dot{N}_g\). In addition, given the level of generality of Eqs.~\eqref{eq:dominant-open-system-correction} and \eqref{eq:diag-1-0-density-matrix}, we expect that a similar perturbation occurs in the models considered in Refs.~\citep{Fatemi:2021wm,Peyruchat:2021tb}.

\section{DC current measurement\label{sec:DC-current-measurement}}

\begin{figure}
	\centering
	\includegraphics[width=\linewidth]{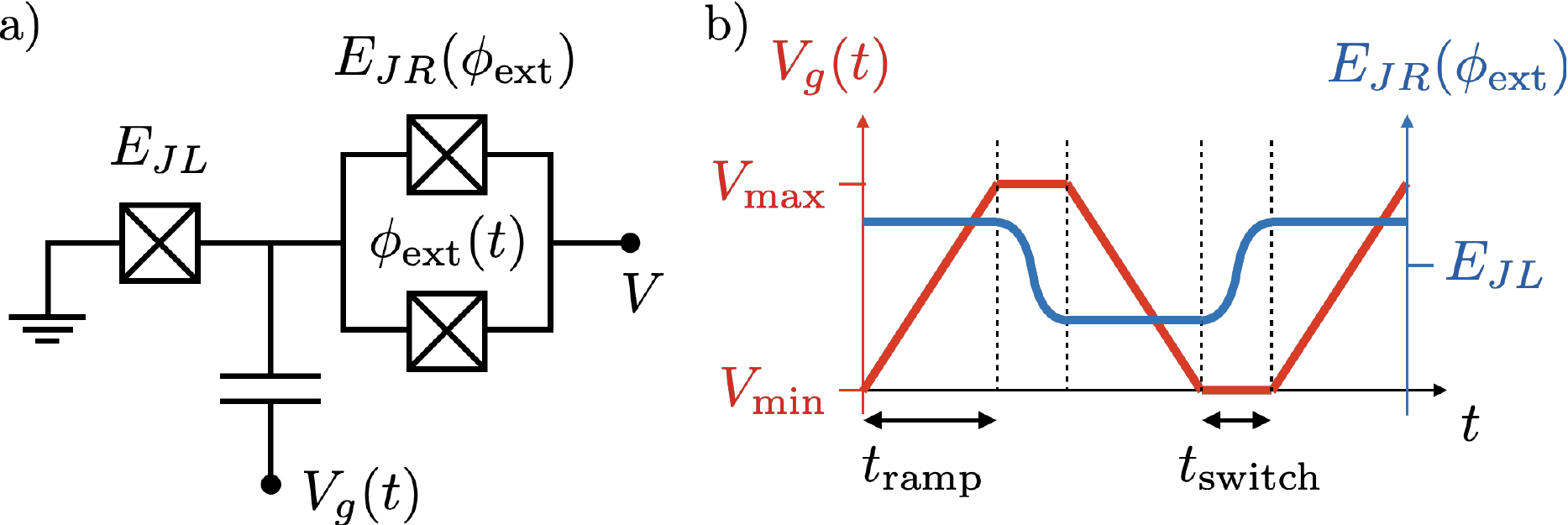}
	
	\caption{Measurement protocol.
		a) The right Josephson junctions is replaced by a SQUID whose energy $E_{JR}\left(\phi_{\text{ext}}\right)$ can be tuned by the magnetic flux $\phi_{\text{ext}}$. The voltage $V$ drives the phase difference continuously, while the linearly time-dependent gate voltage $V_{g}\left(t\right)$ induces a dc current $I_{\text{ind}}\propto\dot{V}_{g}$.
		b) A suggestion on how to tune $N_{g}$ via $V_{g}$ and $E_{JR}$ via $\phi_{\text{ext}}$ as a function of time in order to be able to measure a quantized dc current. Whenever we ramp up $N_{g}$, the induced current flows into the right lead and when we ramp it back down, the current flows from the left lead into the system. In the intermediate steps, $N_{g}$ is held constant to adjust $\phi_{\text{ext}}$ such that $E_{JR}$ becomes larger or smaller than $E_{JL}$.}
	\label{fig:Measurement-protocol}
\end{figure}

As we have indicated in Sec.~\ref{sec:Quantized-current-response},
a remaining experimental obstacle concerns the fact that $N_{g}$
cannot be ramped up indefinitely since at some point the transistor
will break. This limitation can easily be circumvented with a simple
procedure using the fact that the direction of the quantized dc-current
is sensitive to the  junction asymmetry. Namely,
we replace the right junction with a superconducting quantum interference
device (SQUID) consisting of two parallel junctions, each with an
energy $E_{JS} > E_{JL} / 2$ (see Fig.~\ref{fig:Measurement-protocol}a).
This introduces a tunable Josephson energy $E_{JR}\rightarrow E_{JR}\left(\phi_{\text{ext}}\right)=2E_{JS}\cos\left(\phi_{\text{ext}}/2\right)$,
controlled by an external magnetic flux going through the SQUID~\footnote{Here,
we have described a symmetric SQUID. Importantly, however, this
symmetry is not at all required for the proper functioning of this
protocol. All the SQUID needs to fulfill is that there are two regimes,
one where $E_{JR}\left(\phi_{\text{ext}}\right)<E_{JL}$ and one where
$E_{JR}\left(\phi_{\text{ext}}\right)>E_{JL}$.}.

In a first step, the protocol now simply consists of ramping up $N_{g}$ to a maximal value, related to the maximum gate voltage \(V_{\text{max}}\), in one configuration,
e.g., where $E_{JR}\left(\phi_{\text{ext}}\right)>E_{JL}$ (such that the
induced dc current flows into the right contact). Afterwards
$\phi_{\text{ext}}$ is changed to a value where $E_{JR}\left(\phi_{\text{ext}}\right) < E_{JL}$,
and $N_{g}$ is subsequently ramped back down to a minimal value related to the gate voltage \(V_{\text{min}}\) (pumping the dc current into
the system from the left contact), while keeping the bias voltage $V$
on for all times (see Fig.~\ref{fig:Measurement-protocol}b). The time required
to switch the junction asymmetry is referred to as $t_{\text{switch}}$,
while a single ramping goes on for $t_{\text{ramp}}$. For long times
we will measure the averaged dc current 
\begin{equation}
I_{\text{dc}} = \left(1 - \frac{t_{\text{switch}}}{t_{\text{ramp}} + t_{\text{switch}}}\right) e \dot{N}_{g}\,,
\end{equation}
which thus depends only on the single driving parameter $\dot{N}_{g}$
and the two relevant times of the cycle, which are completely controlled
by the experimenter. In the limit of $t_{\text{switch}}\ll t_{\text{ramp}}$,
we find $I_{\text{dc}}=e\dot{N}_{g}$. Note that in a single ramping
process the current $2e\dot{N}_{g}$ flows, as per Eq.~\eqref{eq:IR_dc}
and Eq.~\eqref{eq:IL_dc}. However, we need two individual ramping
processes (ramping the offset charge in both directions) to complete
the cycle, which takes twice the time. Furthermore, we stress that
as long as a transition from $E_{JR}\left(\phi_{\text{ext}}\right)<E_{JL}$
to $E_{JR}\left(\phi_{\text{ext}}\right)>E_{JL}$ can be achieved,
the flux control does not even need to be very precise, nor is it
susceptible to flux noise, apart from the above discussed finite-frequency
perturbations. Note that the only important
restriction is to make sure that \(V_{\text{min}}\) and \(V_{\text{max}}\) are chosen such that the system does not go through the degeneracy point (Weyl point) when ramping from \(E_{JR} < E_{JL}\) to \(E_{JR} > E_{JL}\) (and vice versa). However, we expect even such a dissipation-induced deviation will likely not be dramatic due to this event not being very probable and giving only a small contribution compared to the topological part as long as the difference \(V_{\text{max}} - V_{\text{min}}\) is chosen sufficiently large.

In fact, this protocol has a lot of similarity with Cooper-pair sluices~\citep{Gasparinetti2012,Pekola2013,Niskanen2003,Vartiainen:2007aa}.
However, one advantage of our approach is that we do not require a
precise control of the tunnel couplings to the contacts: the quantization
of the current requires merely the averaging in the $\left(N_{g},\phi\right)$-space,
which is guaranteed in the presence of the bias voltage $V$. Moreover,
contrary to regular Cooper-pair pumps~\citep{Pekola2013}, our proposal
is insensitive to fermion parity, as we argued above.

\section{Conclusion}

We have found that the Cooper-pair transistor hosts topologically
nontrivial Chern numbers, giving rise to a quantization of the dc
current response, which is precisely steered either to the left or
right contact. This circuit has various advantages to alternative
systems, not least the simplicity and straightforward realizability
of the circuit. Surprisingly, low-frequency charge noise is actually
beneficial for the observation of the quantization effect. Moreover,
the Chern number is insensitive to quasiparticle poisoning and to
whether or not the system is in its ground state. The latter is due
to the emergence of Weyl points with higher topological charges connecting
higher bands. Remaining environment-introduced perturbations are found
to be small. Finally, we presented
an experimentally feasible protocol to carry out the dc current measurement.
We conclude that the Cooper-pair transistor presents a promising platform
to realize a topological circuit, with a topological number defined
in a ``mixed'' basis consisting of the phase difference $\phi$
and the offset charge $N_{g}$.

Finally, when considering topological systems, the question of the existence of protected edge states inevitably occurs. Usually, in materials with a topological band structure in $k$-space, edge states naturally occur at the termination of the material (in position space). Since we here consider topological numbers in an alternative base space, the physical presence of edge states becomes somewhat elusive. Nonetheless, they are present in the following sense. A sharp boundary in $x$-space corresponds to a highly nonlocal feature in (canonically conjugate) $k$-space as per the Heisenberg uncertainty principle, such that the edge is able to ``probe'' the bulk topological number directly~\citep{Diehl2021}. In our proposal on the other hand, the Chern number is probed via time-dependent driving, and subsequent time-averaging of the electric current response. Since the current corresponds to the total number of transported particles (per time), and this particle number being conjugate to $\phi$, one can interpret the quantization of the former as an indirect probe of an edge state. The explicit creation of edge states in charge space (e.g., via an active shaping of the charge space itself) will be a crucial future research endeavor, in order to use the above topological features for protected quantum information processing.

\begin{acknowledgments}
We acknowledge fruitful discussions with I. Pop, G. Catelani, V. Fatemi,
L. Bretheau, A. Akhmerov, S. Diehl and P. Bushev. This work has been funded by the German
Federal Ministry of Education and Research within the funding program
Photonic Research Germany under the contract number 13N14891.
\end{acknowledgments}

\appendix

\section{Derivation of topological charges \label{sec:App-Derivation-of-topological-charges}}

\subsection{Single Weyl point\label{sec:App-Single-Weyl-point}}
We here provide the derivation of the Hamiltonian close to the crossing
point between bands $n=0$ and $n=1$, representing a Weyl point with
topological charge $\mathcal{C} = +1$, as in Eq.~\eqref{eq:single_weyl}.

When tuning the Hamiltonian [see Eq.~\eqref{eq:Hamiltonian}] near the degeneracy, that is $N_{g} = - N + 1/2 + \delta N_{g}$,
$E_{JR} = E_{JL} + \delta E_{J}$, and $\phi_{R} = \pi + \delta\phi$ while
at the same time $\phi_{L} = 0$, we find up to first order in \(\delta N_{g}\),  \(\delta E_{J}\), and  \(\delta\phi\)
\begin{align}
	\widehat{H} =  & \frac{E_{C}}{2}\left(\widehat{N}-N+\frac{1}{2}+\delta N_{g}\right)^{2} - E_{JL} \cos\left(\widehat{\varphi}\right) \nonumber \\
	& + (E_{JL} + \delta E_{J}) \cos\left(\widehat{\varphi} - \delta\phi\right) \nonumber \\
	\approx  & \frac{E_{C}}{2}\left(\widehat{N}-N+\frac{1}{2}\right)^{2} + E_{C} \left(\widehat{N}-N+\frac{1}{2}\right) \delta N_{g} \nonumber\\
	& + \delta E_{J} \cos\left(\widehat{\varphi}\right) + E_{JL} \sin\left(\widehat{\varphi}\right) \delta\phi\,.
\end{align}
We now write this Hamiltonian in the charge eigenspace and keep only the relevant subspace involved in the crossing, $\left\{\left|N-1\right\rangle , \left|N\right\rangle \right\}$. In this subspace, we can write the operators as $\widehat{N} = N \left|N\right\rangle\left\langle N\right| + (N-1) \left|N-1\right\rangle\left\langle N-1\right|$ and $\mathrm{e}^{i\widehat{\varphi}} = \left|N\right\rangle\left\langle N-1\right|$, leading to the Hamiltonian
\begin{align}
	\widehat{H} = & \frac{E_{C}}{8}\left(\left|N\right\rangle \left\langle N\right| + \left|N-1\right\rangle \left\langle N-1\right|\right) \nonumber\\
	& + \frac{E_{C}}{2}\delta N_{g} \left(\left|N\right\rangle \left\langle N\right| - \left|N-1\right\rangle \left\langle N-1\right|\right) \nonumber\\
	& + \frac{\delta E_{J}}{2} \left(\left|N\right\rangle\left\langle N-1\right| + \left|N-1\right\rangle\left\langle N\right|\right) \nonumber\\
	& + \frac{E_{JL}}{2} \delta\phi \left(-i \left|N\right\rangle\left\langle N-1\right| + i \left|N-1\right\rangle\left\langle N\right|\right) \nonumber\\
	= & \frac{E_{C}}{8}\, \widehat{\mathbb{I}} + \frac{E_{C}}{2}\delta N_{g}\, \widehat{\sigma}_{z} + \frac{\delta E_{J}}{2}\, \widehat{\sigma}_{x} + \frac{E_{JL}}{2} \delta\phi\, \widehat{\sigma}_{y}\,,
\end{align}
where \(\widehat{\mathbb{I}} = \left|N\right\rangle \left\langle N\right| + \left|N-1\right\rangle \left\langle N-1\right|\) and \(\widehat{\sigma}_{x}\), \(\widehat{\sigma}_{y}\), \(\widehat{\sigma}_{z}\) are the usual Pauli matrices with \(\widehat{\sigma}_{z} = \left|N\right\rangle \left\langle N\right| - \left|N-1\right\rangle \left\langle N-1\right|\).
Note that we can ignore the first term, because, within the subspace spanned by the states \(\vert N\rangle\) and \(\vert N-1\rangle\), it can be regarded as a constant energy contribution.

\subsection{Double Weyl point \label{sec:App-Double-Weyl-point}}
Now, we derive
 the Hamiltonian near the crossing
point between bands $n=1$ and $n=2$, describing a Weyl point with
topological charge $\mathcal{C} = +2$, as in Eq.~\eqref{eq:double_weyl}.

As pointed out in the main text, we have
to tune to $N_{g}=-N+\delta N_{g}$, $E_{JR} = E_{JL} + \delta E_{J}$, and $\phi_{R} - \phi_{L} = \pi + \delta\phi$ to get close to the double Weyl point.
Here, a gapping can occur only by changing between charge states $\left|N-1\right\rangle $
and $\left|N+1\right\rangle $, which is achieved by a higher-order
process involving the tunneling of two Cooper-pairs via virtual charge
states. We tackle this problem by means of a Schrieffer-Wolff transformation.
First, we are shifting the reference point of the energy by the average
value of the charge subspace \{$\left|N-1\right\rangle $, $\left|N+1\right\rangle $\},
$\widehat{H} \rightarrow \widehat{H} - \frac{1}{2} \operatorname{tr} \bigl[ \widehat{H} \widehat{P} \bigr] $,
with the projector onto this subspace $\widehat{P} = \left|N - 1\right\rangle \left\langle N - 1\right| + \left|N + 1\right\rangle \left\langle N + 1\right|$.
Afterwards, we can write the effective Hamiltonian in the low-energy
regime approximately as
\begin{equation}
	\widehat{H}_{2} = \widehat{P} \widehat{H}_{0} \widehat{P} - \widehat{P} \widehat{V} \left(1-\widehat{P}\right) \frac{1}{\widehat{H}_{0}} \widehat{V} \widehat{P}\,, \label{eq:Schrieffer-Wolff}
\end{equation}
when decomposing the Hamiltonian according to $\widehat{H}=\widehat{H}_{0}+\widehat{V}$,
with
\begin{align}
	\widehat{H}_{0} =& \frac{E_{C}}{2} \sum_{N^\prime} \left[\left(N^\prime - N + \delta N_{g}\right)^{2} - 1\right] \left|N^\prime\right\rangle \left\langle N^\prime\right| \nonumber\\
	\approx& \frac{E_{C}}{2} \sum_{N^\prime} \Bigl[\left(N^\prime - N\right)^2 - 1 + 2 (N^\prime - N) \delta N_{g}\Bigr] \left|N^\prime\right\rangle \left\langle N^\prime\right|,
\end{align}
and
\begin{align}
	\widehat{V} =& \sum_{N^\prime} \biggl[\left(\frac{E_{JL} + \delta E_{J}}{2} \mathrm{e}^{i\delta\phi} - \frac{E_{JL}}{2}\right) \left|N^\prime - 1\right\rangle \left\langle N^\prime\right| + \text{h.c.}\biggr] \nonumber\\
	\approx& \sum_{N^\prime} \biggl[\left(\frac{\delta E_{J}}{2} + i \frac{E_{JL}}{2} \delta\phi\right) \left|N^\prime - 1\right\rangle \left\langle N^\prime\right| + \text{h.c.}\biggr],
\end{align}
keeping only the lowest order in $\delta N_{g}$, $\delta\phi$, and $\delta E_{J}$.
Inserting those two into Eq.~\eqref{eq:Schrieffer-Wolff}, we find
\begin{align}
	\widehat{P}\widehat{H}_0\widehat{P} = E_{C} \delta N_{g} \left(\left|N+1\right\rangle \left\langle N+1\right| - \left|N-1\right\rangle \left\langle N-1\right|\right), \label{eq:schrieffer-wolff_zero-order}
\end{align}
and
\begin{align}
	\widehat{P} \widehat{V} \frac{1-\widehat{P}}{\widehat{H}_{0}} \widehat{V} \widehat{P} =& \left(\frac{|v|^2}{\epsilon_{N}} + \frac{|v|^2}{\epsilon_{N+2}}\right) \left|N+1\right\rangle \left\langle N+1\right| \nonumber\\
	&+ \left(\frac{|v|^2}{\epsilon_{N}} + \frac{|v|^2}{\epsilon_{N-2}}\right) \left|N-1\right\rangle \left\langle N-1\right| \nonumber\\
	&+ \frac{v^2}{\epsilon_{N}} \left|N-1\right\rangle \left\langle N+1\right| + \text{h.c.}\,,\label{eq:schrieffer-wolff_correction}
\end{align}
where we defined $\epsilon_{N^{\prime}} \equiv E_{C}/2\, \bigl[\left(N^{\prime} - N\right)^2 - 1\bigr]$ and $v \equiv (\delta E_{J} + i E_{JL} \delta\phi) / 2$. We ignore the first two terms of Eq.~\eqref{eq:schrieffer-wolff_correction} since they give us only what can be regarded as constant energy contributions. Inserting Eqs.~\eqref{eq:schrieffer-wolff_zero-order} and \eqref{eq:schrieffer-wolff_correction} back into Eq.~\eqref{eq:Schrieffer-Wolff}, we find
\begin{equation}
	\widehat{H}_2 = E_{C} \delta N_{g}\, \widehat{\sigma}_{z} + \frac{\delta E_{J}^2 - E_{JL}^2 \delta\phi^2}{2 E_C}\, \widehat{\sigma}_{x} + \frac{E_{JL} \delta E_{J} \delta\phi}{E_C}\, \widehat{\sigma}_{y}\,,
\end{equation}
arriving at Eq.~\eqref{eq:double_weyl}.

\section{Open-system correction terms \label{sec:App-Open-system-corrections}}

Here, we apply perturbation theory to the master equation, Eq.~\eqref{eq:master-eq-instantaneous}, to derive the corrections to the current expectation value as shown in Eq.~\eqref{eq:dc-current_open}, arising through a coupling to the environment and a time-dependent driving of the system.
We vectorize $\widehat{\rho}$ and decompose it into diagonal and off-diagonal sectors (in the eigenbasis of $\widehat{H}$) $ \left\vert \rho \right) = \left( \left\vert \rho_{d} \right), \left\vert \rho_{o} \right) \right) $, each of which are represented by vectors in the Fock-Liouville space
\begin{equation}
	\left\vert \rho_{d} \right) = 
	\begin{pmatrix}
		\rho_{00}\\
		\rho_{11}\\
		\rho_{22}\\
		\vdots
	\end{pmatrix},\qquad
	\left\vert \rho_{o} \right) = 
	\begin{pmatrix}
		\rho_{01}\\
		\rho_{10}\\
		\rho_{02}\\
		\vdots
	\end{pmatrix}.
\end{equation}
In the same manner, we write the superoperators in the corresponding matrix representation with four subblocks.
The two diagonal subblocks of these matrices correspond to transitions from diagonal to diagonal, respectively off-diagonal to off-diagonal sectors. The off-diagonal matrix subblocks describe the coupling between the diagonal and off-diagonal sectors, such that the master equation has the following shape
\begin{multline}
	\left(\begin{matrix}
		\vert \dot{\rho}_{d})\\
		\vert \dot{\rho}_{o})
	\end{matrix}\right)
	= \left[-i
	\left(\begin{matrix}
		0 & \delta\mathbf{L}_{do}\\
		\delta\mathbf{L}_{od} & \mathbf{L}_{oo} + \delta\mathbf{L}_{oo}
	\end{matrix}\right)\right.\\
	\left.+
	\left(\begin{matrix}
		\mathbf{W}_{dd} & \mathbf{W}_{do}\\
		\mathbf{W}_{od} & \mathbf{W}_{oo}
	\end{matrix}\right)
	\right]
	\left(\begin{matrix}
		\vert \rho_{d})\\
		\vert \rho_{o})
	\end{matrix}\right), \label{eq:master-eq-in-sectors}
\end{multline}
where $ \mathbf{L}_{oo} $ denotes the nonzero part of the Liouvillian superoperator $ \mathbf{L}_{0} $.

We demand that $\mathbf{W}_{dd}$ has one (nondegenerate) zero eigenvalue (equivalent to it having a stationary state) $\mathbf{W}_{dd} \bigl\vert \rho_{d}^{(0, 0)} \bigr) = 0$. Let us additionally suppose that $\mathbf{W}_{dd}$ relaxes the system to the ground state (up to exponentially suppressed contributions, equivalent to a low temperature assumption) such that $\bigl\vert \rho_{d}^{(0, 0)} \bigr) = \vert 0_{d})$ (the vectorized version of \(\widehat{\rho}^{(0,0)} = |0\rangle \langle 0|\)  appearing in the main text as \(\widehat{\rho}_{0}\)). The subscript $d$ simply expresses the fact that here, the size of the vector is that of the diagonal sector, such that
\begin{equation}
	\bigl\vert \rho^{(0,0)} \bigr) = \left| 0 \right) =
	\begin{pmatrix}
		\bigl\vert 0_{d} \bigr)\\
		0
	\end{pmatrix}.
\end{equation}
Be aware that this state is not the exact stationary state of the total kernel $\mathbf{W}$ because we are explicitly allowing transitions between diagonal and off-diagonal states in the kernel, represented by the off-diagonal subblocks. Corrections to $ \widehat{\rho}^{(0,0)} $ are deviations from the commonly assumed rotating-wave approximation (RWA). Therefore, in the absence of the time-dependent driving, we will always end up in a stationary state of the shape $\widehat{\rho}^{\text{st}} = \widehat{\rho}^{(0,0)} + \widehat{\rho}^{(0,1)}$, where $\widehat{\rho}^{(0,1)}$ denotes the first-order correction beyond  RWA. The notation $(0,1)$ refers to this contribution being of zero order in the driving $\left\Vert \delta\mathbf{L}\right\Vert$ and of linear order in the kernel $\left\Vert \mathbf{W}\right\Vert$; see also Eq.~\eqref{eq:expanded_density_matrix}. However, the stationary state, \(\widehat{\rho}^{\text{st}}\), does not contribute to the dc part of the current. As we already argued in the main text, this is due to the fact that dc currents can be induced only by driving the system out of equilibrium and therefore cannot be connected to terms of zero order in the driving with the environment at equilibrium~\citep{footnote02}. 

Turning on the slow driving of the system now gives us a small correction to the density matrix, driving the system away from the stationary state to what we call a quasistationary state,  $\widehat{\rho}$. Assuming that $\left\Vert \delta\mathbf{L}\right\Vert <\left\Vert \mathbf{W}\right\Vert <\left\Vert \mathbf{L}_{0}\right\Vert $, we can expand this state in both $\left\Vert \mathbf{W}\right\Vert$ and $\left\Vert \delta\mathbf{L}\right\Vert$, analogously to Eq.~\eqref{eq:expanded_density_matrix}, as
\begin{equation}
	\left\vert \rho \right) = \sum_{\nu, \mu}\, \bigl\vert \rho^{(\nu, \mu)} \bigr)\,. \label{eq:expanded_density_vector}
\end{equation}
We now define the quasistationary state $\widehat{\rho}$ such that the time derivative of each order $\partial_t \bigl\vert \rho^{(\nu, \mu)}\bigr)$ is of higher order in the driving \((\nu+1, \mu)\). Here, the zeroth order, \(\bigl\vert \rho^{(0,0)} \bigr) = \left| 0 \right)\), is constant, meaning it only implicitly depends on time due to the now time-dependent basis (which is the instantaneous eigenbasis of $\widehat{H}$).

Besides the previously mentioned \(\bigl\vert \rho^{(0,1)} \bigr)\), we now find the first-order correction \(\bigl\vert \rho^{(1,0)} \bigr)\) which has a nonzero dc contribution and therefore gives rise to the leading-order terms of the dc current. Even though \(\bigl\vert \rho^{(0,1)} \bigr)\) does not contribute to the dc current, we
nonetheless have to compute its off-diagonal part $ \bigl\vert \rho_{o}^{(0,1)} \bigr) $ because it is a necessary intermediate result to calculate the diagonal sector of the correction $ \bigl\vert \rho^{(1,0)} \bigr) $.

Reexpressing Eq.~\eqref{eq:master-eq-in-sectors} for the diagonal and off-diagonal sectors separately
to find the quasistationary state, we find
\begin{align}
	\partial_t \bigl\vert \rho_{d}\bigr) =& \mathbf{W}_{dd} \vert \rho_{d}) + \left[-i\delta\mathbf{L}_{do} + \mathbf{W}_{do}\right] \vert \rho_{o})\,,
	\label{eq:diag-master-eq}\\
	\partial_t \bigl\vert \rho_{o}\bigr) =& \left[-i\delta\mathbf{L}_{od} + \mathbf{W}_{od}\right] \vert \rho_{d})\nonumber\\
	&- \left[i (\mathbf{L}_{oo} + \delta\mathbf{L}_{oo}) - \mathbf{W}_{oo}\right] \vert \rho_{o}).
	\label{eq:off-diag-master-eq}
\end{align}
We begin by considering the stationary open system (i.e., in absence of the driving) equivalent to keeping only terms of zero order in the driving. In leading order of $\left\Vert \mathbf{W}\right\Vert$, we find
\begin{align}
	0&= \mathbf{W}_{dd} \bigl\vert \rho_{d}^{(0,1)} \bigr) + \mathbf{W}_{do} \bigl\vert \rho_{o}^{(0,1)} \bigr)\,, \label{eq:diag-0,2-master-eq}\\
	0&= \mathbf{W}_{od} \bigl\vert 0_{d} \bigr) - i\mathbf{L}_{oo} \bigl\vert \rho_{o}^{(0,1)} \bigr)\,, \label{eq:off-diag-0,1-master-eq}
\end{align}
providing the stationary state
\begin{equation}
	\bigl\vert \rho^{\text{st}} \bigr) =
	\begin{pmatrix}
		\bigl\vert 0_{d} \bigr) - \mathbf{\widetilde{W}}_{dd}^{-1} \mathbf{W}_{do} \bigl\vert \rho_{o}^{(0,1)} \bigr)\\
		\bigl\vert \rho_{o}^{(0,1)} \bigr)
	\end{pmatrix},
\end{equation}
with
\begin{equation}
	\bigl\vert \rho_{o}^{(0,1)} \bigr) = -i \frac{1}{\mathbf{L}_{oo}} \mathbf{W}_{od} \bigl\vert 0_{d} \bigr)\,.
\end{equation}
Here, $\mathbf{\widetilde{W}}_{dd}^{-1}$ is defined such that
\begin{equation}
	\mathbf{\widetilde{W}}_{dd}^{-1} \mathbf{W}_{dd} = \mathbf{I}_{dd} - \bigl|0_{d}\bigr) \bigl(0_{d}\bigr|,
\end{equation}
with $\mathbf{I}_{dd}$ being the identity matrix in the diagonal subspace and $(0_{d}|$ being the left eigenvector of $\mathbf{W}_{dd}$ with eigenvalue zero, which is the trace $\left(0\right|\bullet = \operatorname{tr} [\bullet]$. We use this notation to denote a map from an operator to a scalar, such that the scalar product $ (0 |0) = \operatorname{tr} \bigl[ \widehat{\rho}^{(0,0)} \bigr] = 1 $.

We can derive the corrections for the quasistationary state in the presence of the drive by comparing the terms of Eqs.~\eqref{eq:diag-master-eq} and \eqref{eq:off-diag-master-eq} that are of linear order in \(\left\Vert \delta\mathbf{L}\right\Vert\), where we find in leading order of \(\left\Vert \mathbf{W}\right\Vert\)
\begin{align}
	\partial_t \bigl\vert \rho_{d}^{(0,1)}\bigr) &= \mathbf{W}_{dd} \bigl\vert \rho_{d}^{(1,0)} \bigr) - i\delta\mathbf{L}_{do} \bigl\vert \rho_{o}^{(0,1)} \bigr) + \mathbf{W}_{do} \bigl\vert \rho_{o}^{(1,0)} \bigr)\,, \label{eq:diag-1,1-master-eq}\\
	0 &= -i\delta\mathbf{L}_{od} \bigl\vert 0_{d} \bigr) - i\mathbf{L}_{oo} \bigl\vert \rho_{o}^{(1,0)} \bigr)\,. \label{eq:off-diag-1,0-master-eq}
\end{align}
This second equation is fulfilled for
\begin{equation}
	\bigl\vert \rho_{o}^{(1,0)} \bigr) = -\frac{1}{\mathbf{L}_{oo}} \delta\mathbf{L}_{od} \bigl\vert 0_{d} \bigr)\,,
\end{equation}
which directly gives us  the Thouless result~\citep{Thouless1983} [see Eq.~\eqref{eq:adiabatic_wave}],
\(\bigl\vert \rho_{d}^{(0,0)} \bigr) + \bigl\vert \rho_{o}^{(1,0)} \bigr)\). This solution now receives an additional correction in the diagonal sector which one can infer from Eq.~\eqref{eq:diag-1,1-master-eq},
\begin{multline}
	\bigl\vert \rho_{d}^{(1,0)} \bigr) = \left[\mathbf{\widetilde{W}}_{dd}^{-1} \left(\delta \mathbf{L}_{do} \frac{1}{\mathbf{L}_{oo}} \mathbf{W}_{od} + \mathbf{W}_{do} \frac{1}{\mathbf{L}_{oo}} \delta \mathbf{L}_{od}\right.\right.\\
	\left.\left.+ i\partial_{t} \Bigl[\widetilde{\mathbf{W}}_{dd}^{-1} \mathbf{W}_{do} \frac{1}{\mathbf{L}_{oo}} \mathbf{W}_{od}\Bigr]\right) \bigl\vert 0_{d} \bigr)\right].
	\label{eq:diag-leading-order-rho-corr}
\end{multline}
We stress that this additional correction arises solely because of the nonzero off-diagonal subblock in the kernel, \(\mathbf{W}_{do}\), which would vanish when making the RWA. Strikingly, it is thus an open-system correction of zero-order in \(\left\Vert \mathbf{W}\right\Vert\). This contribution, Eq.~\eqref{eq:diag-leading-order-rho-corr}, gives rise to the leading-order open-system correction to the current; see Eq.~\eqref{eq:dominant-open-system-correction}.

For a system-reservoir interaction that conserves the current (see discussion in the main text), we can write the current expectation value as $ I_{\alpha} = \operatorname{tr} \bigl[ \widehat{I}_{\alpha} \widehat{\rho} \bigr] = \left( I_{\alpha} \vert \rho \right) $, introducing the notation $ \left( A \right\vert \bullet = \operatorname{tr} \bigl[ \widehat{A}\, \bullet \bigr] $. $I_{\alpha}$ can be computed with the help of the operator $\widehat{N}_{\alpha} = -i \sum_{n, m} \left\langle n\right| \partial_{\phi_{\alpha}} \left| m \right\rangle \left| n \right\rangle \left\langle m \right|$, as already introduced in Eq.~\eqref{eq:current_1,0-contribution_Berry-curvature-term}.
It can easily be shown that
\begin{equation}
	\widehat{I}_{\alpha} = 2e \sum_{n} \partial_{\phi_\alpha} \epsilon_n \left| n \right\rangle \left\langle n \right| + 2e\, i\Bigl[ \widehat{N}_{\alpha}, \widehat{H} \Bigr]\,,
\end{equation}
from which directly follows
\begin{equation}
	I_{\alpha}
	= 2e \sum_{n} \rho_{nn}\, \partial_{\phi_\alpha} \epsilon_n + 2e \left( N_{\alpha} \right\vert i\mathbf{L}_0 \left\vert \rho \right),
\end{equation}
which we have written in the form $ I_{\alpha} = I_{\alpha, d} + I_{\alpha, o} $, analogously to the current in the closed system in Eq.~\eqref{eq:current_of_time}. Here, \(I_{\alpha, d}\) (\(I_{\alpha, o}\)) represents the current contribution due to the diagonal (off-diagonal) elements of the density matrix.
The diagonal term, \( I_{\alpha, d} = \sum_{n} \rho_{nn}\, I_{\alpha, n}^{(0)}\),
represents a generalized version of the zero-order contribution corresponding to the ordinary Josephson effect of the closed system, $ I_{\alpha, n}^{(0)} = 2e\, \partial_{\phi_\alpha} \epsilon_n $ [see Eq.~\eqref{eq:current_of_time}].
However, when inserting the order-by-order expansion of the density matrix, Eq.~\eqref{eq:expanded_density_vector}, we find that in contrast to the closed-system result, this expression contributes to the dc current, in leading-order via \(I_{\alpha, d}^{(1, 0)} = 2e \sum_{n} \rho_{nn}^{(1, 0)}\, \partial_{\phi_\alpha} \epsilon_n\), with \(\rho_{d}^{(1, 0)}\) as in Eq.~\eqref{eq:diag-leading-order-rho-corr}. The terms of zeroth order in the driving \(I_{\alpha, d}^{(0, \mu)}\), on the other hand, cannot create a dc contribution with the environment in equilibrium, as discussed above.

Similarly, we can expand the off-diagonal term, \(I_{\alpha, o} = \sum_{\nu, \mu} I_{\alpha, o}^{(\nu, \mu)}\), and find that the leading-order term with nonvanishing dc part is first order in the driving and giving rise to the Berry curvature, $ I_{\alpha, o}^{(1, 0)} = I_{\alpha, 0}^{(1)} = - 2e \left( N_{\alpha} \right\vert i\delta\mathbf{L} \left\vert 0 \right)$; see Eq.~\eqref{eq:current_of_time}.
The linear open-system correction $ I_{\alpha, o}^{(0, 1)} = 2e \left( N_{\alpha} \right\vert \mathbf{W} \left\vert 0 \right)$
is again purely contributing to the ac current, like any higher-order term that does not depend on the driving parameters.
We thus find the dc current response \(\overline{I}_\alpha = \overline{I}_{\alpha, o}^{(1,0)} + \overline{I}_{\alpha, d}^{(1,0)}\) as in Eq.~\eqref{eq:dc-current_open}, where \(\overline{I}_{\alpha, o}^{(1,0)} = \overline{I}_{\alpha, 0} = -2e \dot{N}_g C_{\alpha,0}\) is the topologically quantized result for the closed system and \(\overline{I}_{\alpha, d}^{(1,0)} = 2e \sum_{n} \rho_{nn}^{(1, 0)}\, \partial_{\phi_\alpha} \epsilon_n\) is the leading-order correction due to the open-system description.

\bibliography{paper1}

\begin{thebibliography}{142}
\bibitem [{foo()}]{footnote02}%
  \BibitemOpen
  \href@noop {} {\bibinfo {title} {This is due to all the \(\hat{\rho}^{(0,
  \mu)}\) being symmetric in \(\phi\) while the operator \(\hat{I}_\alpha\) is
  antisymmetric. therefore, an integral over an entire period of \(\phi\) has
  to vanish.}}\BibitemShut {Stop}%
\end{thebibliography}

\end{document}